\documentclass[usenatbib]{mn2e}
\usepackage{graphicx}
\usepackage{float}
\usepackage{epsfig}
\usepackage{amsmath} 
\usepackage{tabu}
\usepackage{subfig}
\usepackage{color,soul}

\title{CO line ratios in molecular clouds: the impact of environment }
\author[C.H.Pe\~naloza et al.]{Camilo H. Pe\~naloza$^{1}$, Paul C. Clark$^{1}$, Simon C.O. Glover$^{2}$,
\and \& Ralf S. Klessen$^{2,3}$ \\
$^{1}$School of Physics and Astronomy, The Queens Buildings, The Parade, Cardiff University, Cardiff CF24 3AA, UK\\
$^{2}$Universit\"{a}t Heidelberg, Zentrum f\"{u}r Astronomie, Institut f\"ur Theoretische Astrophysik, Albert-Ueberle-Str. 2, 69120 Heidelberg, Germany \\
$^{3}$Universit\"{a}t Heidelberg, Interdiszipli\"{a}res Zentrum f\"{u}r Wissenschaftliches Rechnen, INF 205, 69120 Heidelberg}

\begin{document}

\maketitle

\begin{abstract}
Line emission is strongly dependent on the local environmental conditions in which the emitting tracers reside. In this work, we focus on modelling the CO emission from simulated giant molecular clouds (GMCs), and study the variations in the resulting line ratios arising from the emission from the  $J=1-0$, $J=2-1$ and $J=3-2$ transitions. We perform a set of smoothed particle hydrodynamics (SPH) simulations with time-dependent chemistry, in which environmental conditions -- including total cloud mass, density, size, velocity dispersion, metallicity, interstellar radiation field (ISRF) and the cosmic ray ionisation rate (CRIR)  -- were systematically varied. The simulations were then post-processed using radiative transfer to produce synthetic emission maps in the 3 transitions quoted above. We find that the cloud-averaged values of the line ratios can vary by up to $\pm 0.3$ dex, triggered by changes in the environmental conditions. Changes in the ISRF and/or in the CRIR have the largest impact on line ratios since they directly affect the abundance, temperature and distribution of CO-rich gas within the clouds. We show that the standard methods used to convert CO emission to H$_2$ column density can underestimate the total H$_2$ molecular gas in GMCs by factors of 2 or 3, depending on the environmental conditions in the clouds.    
\end{abstract}


\section{Introduction} \label{introsec}

The evolution, structure and physical properties of giant molecular clouds (GMCs) are highly dependant on the surrounding environmental conditions. The changes in the surrounding environment of GMCs can have a direct impact on the formation of stars, since it is within these complicated cloud complexes that most of the molecular gas, which eventually will be transformed into stars, is contained \citep{2016SAAS...43...85K}. Gas is mostly found in the form of molecular hydrogen (H$_2$) that due to the low temperatures of GMCs cannot be directly observed. Therefore empirically derived relations help estimate the total molecular content by making use of other molecular tracers.

Carbon monoxide's (CO) easily excited rotational lines make it one of the most well known tracers of the interstellar medium (ISM), as such we rely on CO emission to estimate the total molecular content of a GMC. The X-factor ($X_{\rm CO}$) allows us to estimate the total molecular gas of a GMC from the integrated intensity of the $J=1-0$ emission line of CO by 

\begin{equation}
 N_{\rm H_2} = X_{\rm CO} W_{10},
 \label{eqn:xco}
\end{equation}

\noindent where $N_{\rm H_2}$ is the column density of H$_2$, $W_{10}$ the integrated intensity of the $J=1-0$ transition of CO and $X_{\rm CO}$ the empirically derived conversion factor. Several studies have already explored the accuracy of $X_{\rm CO}$ and its dependence on environmental conditions \citep{Bolatto:2013hl,2012ApJ...751...10P,2013ApJ...777....5S,2014MNRAS.440L..81O, 2015MNRAS.452.2057C,2015ApJ...812....6B}. 

As distances become larger, it becomes harder to resolve far away clouds in nearby galaxies. Extragalactic studies therefore rely on higher rotational transitions with smaller wavelengths since they permit higher-resolution imaging. Most commonly used are the $J=2-1$ and $J=3-2$ rotational transition lines since they are bright and easily observed. The drawback however is that $X_{\rm CO}$ is an empirical relation that is only calibrated for the $J=1-0$ emission line. Therefore, the integrated intensity of $J=2-1$ emission line ($W_{21}$) is converted to $W_{10}$ by using the empirically derived ratio $R_{21} = W_{21} / W_{10}$. Similarly the empirically derived $R_{31} = W_{32} / W_{10}$ is used to convert from $W_{32}$ to $W_{10}$. One then applies the X-factor to convert the CO emission into a column density of H$_2$.

The ratios $R_{21}$ and $R_{31}$ thus play a key role in our understanding of galactic-scale star formation relations, such as the Kennicutt-Schmidt relationship \citep{Schmidt1959, Kennicutt1998}. Typically, extragalactic studies adopt a value of $R_{21}$ = 0.7 \citep{1990ApJ...348..434E,1991A&A...251....1C,1995A&A...303..851B,1997ApJ...486..276S,1997IAUS..170...39H,2001ApJS..136..189S,2008AJ....136.2846B,2009AJ....137.4670L,2011MNRAS.416.1250B} to convert from $W_{21}$ to $W_{10}$. However, this could be inaccurate given the results shown in \citet{2017MNRAS.465.2277P}, which suggest $R_{21}$ has a bimodal distribution dependant on the physical conditions surrounding the emitting gas. Another example of a widely-used ratio is $R_{31}$, which is mostly used to study star formation in high redshift galaxies \citep{2010ApJ...718..177A,2013ApJ...768..132B,2014MNRAS.442..558A,2015A&A...577A..46D}. In most of these cases the $J=3-2$ emission line is observed and then converted using the standard value $R_{31}= 0.5$ \citep{2014MNRAS.442..558A}, before deriving any physical properties of the system. 

Fortunately numerical simulations provide a way through which these ratios can be studied and their behaviour and dependancies on environment properly quantified. In this paper we numerically follow the evolution of GMCs that are post-processed to generate synthetic observations. The aim is to gain a better understanding of the ratios of CO's rotational emission lines and how they are influenced by changes in cloud mass, density, metallicity, the strength of the Interstellar Radiation Field (ISRF) and the cosmic ray ionization rate (CRIR). Therefore we simulate a set of clouds in which the initial conditions are systematically changed in order to cover a wide range of realistic environmental conditions. 

The structure of the paper is as follows. In section \ref{metsec} we describe the numerical setup and the initial conditions used to model the evolution and synthetic observations of these GMCs. In Section \ref{ressec} we present our results. We look at how the cloud's morphology changes depending on environment as well as study the impact this has on the value and distribution of $R_{21}$. In Section \ref{ressec2} we examine how variations in environment impact the observation of unresolved GMCs and the consequences this has on different line ratios. In Section \ref{dissec} we discuss how variations in $R_{21}$ and $R_{31}$ affect calculated column densities of H$_2$ as well as whether $R_{21}$ can trace changes in CO abundances. Finally we summarise our findings in Section \ref{consec}.

\section{Method}\label{metsec}

\subsection{Hydrodynamics and chemistry}

To model the gas in this study, we use a modified version of the publicly available smoothed particle hydrodynamics (SPH) code, GADGET-2 \citep{Springel:2005cz}. These modifications include a time-dependent chemical network that follows the formation and destruction of H$_2$  \citep{Glover:2007gr,Glover:2007wq} and CO \citep{NelsonLanger1999}, more details of which can be found in \citet{Glover:2012et}, which also includes the photodissociation rates that we adopt in this study.  We adopt the same radiative heating and cooling rates, and cosmic ray heating rate as described in \citet{Glover:2007gr} and \citet{Glover:2012dd}. To treat the attenuation of the ISRF we use the {\sc TreeCol} algorithm developed by \citet{Clark_etal_2012}.

\subsection{Initial conditions}
\label{sec:ics}

We produce a set of numerical simulations with different initial conditions to study the impact of environment on the evolution of GMCs and the impact this has on CO emission lines. The initial setup of all the clouds is a uniform sphere where a turbulent velocity field with a power spectrum of $P(k)\propto k^{-4}$ is imposed and left to decay as the cloud evolves. Since the aim of this study is to look at the structure and evolution of GMCs prior to the onset of star formation, we therefore stop each run just before the onset of star formation modelled by the creation of so called sink particles (see \citealt{Glover:2012dd}). It is important to note that since the initial conditions affect the evolution of each cloud, when star formation is triggered will be at different times for each run. We make use of the clouds simulated in previous papers by \citet{2015MNRAS.452.2057C} and \citet{2016MNRAS.456.3596G} since they already cover part of the parameter space we intend to study. Below we cover what the variations in initial conditions are but refer the reader to those papers for full details.

First we summarise the initial conditions of the simulations by \citet{2015MNRAS.452.2057C}. They cover a range of different ISRFs intensities that are scaled proportional to $G_0$, where $G_0 = 1.7$ in \citet{Habing1968} units and a range of CRIRs that are scaled proportional to $\zeta_{\rm H} = 3\times 10^{-17}\ {\rm s^{-1}}$. These clouds have a mass of either $10^4 M_{\odot}$ or $10^5 M_{\odot}$. Additionally the initial density is varied to be either $n = 100\ {\rm cm^{-3}}$ or $n = 10^{4} {\rm cm^{-3}}$. Lastly since the initial state of the gas can delay the formation of CO and therefore its total emission, the initial molecular fraction is changed to be either $f({\rm H_2})=1$ or $f({\rm H_2})=0$, i.e. fully molecular or fully atomic. All of these runs were performed with a turbulent velocity field generated from a `natural' mix of solenoidal and compressive modes in a 2:1 ratio. 

The simulations by \citet{2016MNRAS.456.3596G} have an initial mass of $10^4 M_{\odot}$, initial density of $n=276\ {\rm cm^{-3}}$, have an initial molecular fraction of $f({\rm H_2})=0$ and have a turbulent velocity field that is generated from purely solenoidal modes. In addition the ISRF and CRIR are scaled in the same way as \citet{2015MNRAS.452.2057C}, however the CRIR is scaled proportional to $\zeta_{\rm H} = 1 \times 10^{-17}\ {\rm s^{-1}}$. Finally the metal fraction is varied with respect to solar metallicity ($Z_{\odot}$), adopting values of $Z = Z_{\odot}$, $Z= 0.5\ Z_{\odot}$ and $Z = 0.2\ Z_{\odot}$. 

Taken together, these two sets of simulations cover a wide range of parameter space. However, there are still cases where it is difficult to compare the clouds, as several cloud properties are changing at once.  To isolate the effect of varying individual environmental properties, we thus perform an additional set of simulations for our current study.  First, \citet{2015MNRAS.452.2057C} and \citet{2016MNRAS.456.3596G} scale the ISRF and CRIR {\em together} making it hard to disentangle the effect of either, therefore we run four clouds that vary either the ISRF or the CRIR. Additionally the small mass clouds in \citet{2015MNRAS.452.2057C} and all of the clouds in \citet{2016MNRAS.456.3596G} have a slightly different initially density, different turbulent velocity field, a different $\alpha_{vir}$ and slightly different $\zeta_{\rm H}$. As such we present four extra simulations where only one of these parameters is varied. 

Note that the dynamical state of our clouds is primarily controlled by the ratio of gravitational to (turbulent) kinetic energy. In all our clouds the thermal energy in the gas is insufficient to unbind the cloud, even in the cases where we adopt high ISRFs and high CRIRs.  For more details one can also see the study by \citet{2015MNRAS.451.3679B} where they explore the binding and star-forming properties of cloud that are exposed to even higher ISRFs and CRIRs than we adopt here. In addition, clouds from \citet{2016MNRAS.456.3596G} apply an external pressure term to prevent ``evaporation'' of material \citep{1990nmns.work..269B}. However we find that the external pressure has minimal influence on whether our clouds will collapse and form stars.

We summarise the set of simulations in Table \ref{tab:InitialConditions}. Note that the IDs given in this table will be used throughout the rest of the paper.

\begin{table*}
  \begin{tabular*}{\textwidth}{ cccccccccc }
    \hline
     ID & $n$ & $M$ & $\alpha_{vir}$& ISRF & CRIR & $f({\rm H_2})$ & Z &Turbulence &  Time$^a$ \\ \hline
      & $[\rm cm^{-3}]$ & $[M_{\odot}]$ & &$[G_{0}]$ & $ [10^{-17}\ {\rm s^{-1}}]$ & &$[Z_{\odot}]$& & $[Myr]/[t_{\rm ff}]/[t_{cr}]$ 
    \\ \hline      
     \multicolumn{10}{ c }{\citet{2015MNRAS.452.2057C}} \\ \hline
    CG15-M4-G1 & 100 & $10^4$ & 0.5 & 1 &3 & 1 & 1 & Natural & 1.83/0.42/0.60\\ 
    CG15-M4-G10 & 100 & $10^4$ & 0.5 & 10 & 30 & 1 & 1 & Natural & 2.09/0.48/0.69\\ 
    CG15-M4-G100 & 100 & $10^4$ & 0.5 & 100 & 300 &1 & 1 & Natural & 1.96/0.45/0.65\\ 
    CG15-M5-G1 & 100 & $10^5$ & 0.5 & 1 & 3 & 1 & 1 & Natural & 1.17/0.27/0.39\\ 
    CG15-M5-G10 & 100 & $10^5$ & 0.5 & 10 & 30 & 1 & 1 & Natural & 1.52/0.35/0.50\\ 
    CG15-M5-G100& 100 & $10^5$ & 0.5 & 100 & 300 & 1 & 1 & Natural & 1.39/0.45/0.65\\ 
    CG15-M5-G1/A& 100 & $10^5$ & 0.5 & 1 & 3 &0 & 1 & Natural & 1.31/0.30/0.43\\ 
    CG15-M5-G100/A & 100 & $10^5$ & 0.5 & 100 & 300 & 0 & 1& Natural & 1.26/0.29/0.42\\ 
    CG15-CMZ1 & $10^4$ & $10^5$ & 0.5 & 100 & 300 &1 & 1 & Natural & 0.10/0.23/0.67\\ 
    CG15-CMZ2 & $10^4$ & $10^5$ & 2 & 100 & 300 &1 & 1& Natural & 0.09/0.22/0.32\\ 
    \hline \multicolumn{10}{ c }{\citet{2016MNRAS.456.3596G}} \\ \hline
    GC16-Z1-G1 & 276 & $10^4$ & 1 &1 & 1 & 0 & 1 & Solenoidal & 1.97/0.75/1.53 \\
    GC16-Z1-G10 & 276 & $10^4$ & 1 &10 & 10 & 0 & 1 &Solenoidal & 2.17/0.83/1.68 \\
    GC16-Z1-G100 & 276 & $10^4$ & 1 &100 & 100 & 0 & 1 &Solenoidal & 2.03/0.78/1.57 \\
    GC16-Z05-G1 & 276 & $10^4$ & 1 &1 & 1 & 0 & 0.5 &Solenoidal & 2.44/0.93/1.89 \\
    GC16-Z05-G10 & 276 & $10^4$ & 1 &10 & 10 & 0 & 0.5 &Solenoidal & 2.61/1.00/2.02 \\
    GC16-Z05-G100 & 276 & $10^4$ & 1 &100 & 100 & 0 & 0.5 &Solenoidal & 2.81/1.07/2.18 \\
    GC16-Z02-G1 & 276 & $10^4$ & 1 &1 & 1 & 0 & 0.2 &Solenoidal & 2.79/1.06/2.16 \\
    GC16-Z02-G10 & 276 & $10^4$ & 1 &10 & 10 & 0 & 0.2 &Solenoidal & 2.95/1.13/2.29 \\
    GC16-Z02-G100 & 276 & $10^4$ & 1 &100 & 100 & 0 & 0.2 &Solenoidal & 3.59/1.37/2.78 \\ 
    \hline \multicolumn{10}{ c }{Additional Runs} \\ \hline
    M4-G1-CR30& 100 & $10^4$ & 0.5 & 1 &30 & 1 & 1 & Natural & 2.00/0.46/0.66\\
    M4-G1-CR300 & 100 & $10^4$ & 0.5 & 1 &300 & 1 & 1& Natural & 2.35/0.54/0.77\\
    M4-G10-CR3 & 100 & $10^4$ & 0.5 & 10 &3 & 1 & 1& Natural & 2.09/0.48/0.69\\   
    M4-G100-CR3 & 100 & $10^4$ & 0.5 & 100 &3 & 1 & 1& Natural & 2.13/0.49/0.70\\
    M4-G1-CR1 & 100 & $10^4$ & 0.5 & 1 & 1 & 1 & 1 & Natural & 1.78/0.41/0.59\\ 
    M4-$\alpha$1 & 100 & $10^4$ & 1 & 1 & 3 & 1 & 1& Natural & 1.87/0.43/0.87 \\  
    M4-N300 & 300 & $10^4$ & 0.5 & 1 & 3 & 1 & 1 & Natural & 0.73/0.29/0.42 \\
    M4-SOL & 100 & $10^4$ & 0.5 & 1 & 3 & 1 & 1 & Solenoidal & 3.13/0.72/1.03 \\	\hline
  \end{tabular*}
  \caption{In this table we summarise the initial conditions for each cloud. The virial conditions of the clouds are given by $\alpha_{vir} = {\rm E_{kin} / E_{pot}}$. $G_0$ is given in \citet{Habing1968} units. $f({\rm H_2})$ denotes the initial molecular fraction of the gas and $Z_{\odot}$ its metallicity. \\
$^a$ Run time/fraction of free-fall time/fraction of crossing time ($t_{cr}=R/\langle v \rangle$)}
  \label{tab:InitialConditions}
\end{table*}

\subsection{Post-processing}

Once the hydrodynamical simulation is finished we post-process the snapshot with RADMC-3D \citep{2012ascl.soft02015D} and create synthetic images. Given that we have clouds of different sizes and densities, we used the refining method developed in \citet{2017MNRAS.465.2277P} to account for all particles in the GADGET-2 snapshot. This assures that no information is lost when interpolating particles to the grid and therefore guarantees convergence of the intensity maps which is important when comparing line ratios of different sized clouds. Additionally, we have implemented an extension to the `Sobolev' approximation in RADMC-3D \citep{Shetty:2011eh}, that accounts for both the velocity and density variations within the cloud. By making use of density gradients within the cloud we can better calculate the local optical depth and therefore the total emission for the cloud. A more detailed description is given in Appendix \ref{appsec}. Since radiative transfer simulations of  clouds in \citet{2015MNRAS.452.2057C} and \citet{2016MNRAS.456.3596G} were performed without these additional methods, we re-do the radiative transfer for these clouds. 

For each cloud we create synthetic observations for the first three rotational lines of $^{12}$CO ($J=1-0,J=2-1,J=3-2$). Integrating along the {\it z}-axis i.e. velocity in PPV space, we then create zeroth moment maps for each line. All the final maps have an imposed cut at emissions lower than $0.01\ {\rm K\ km\ s^{-1}}$, this is motivated by our previous study \citep{2017MNRAS.465.2277P}. All these maps are `ideal' synthetic observations since they do not include any noise or telescope effects.

Finally it is worth noting that we only study the first three rotational lines of CO since higher transitions depend on additional physics that are not well-captured by this type of numerical simulation. Higher CO transitions are normally excited by high velocity shocks within the clouds \citep{2013ApJ...779L..19P,2016ApJ...827..107P}; these shocks are not well resolved by 3D numerical simulations, and so the microphysics of such regions are not properly traced by our models.

\section{Results}\label{ressec}

\subsection{Cloud morphology and appearance}

\begin{figure*}
\includegraphics[width=\textwidth]{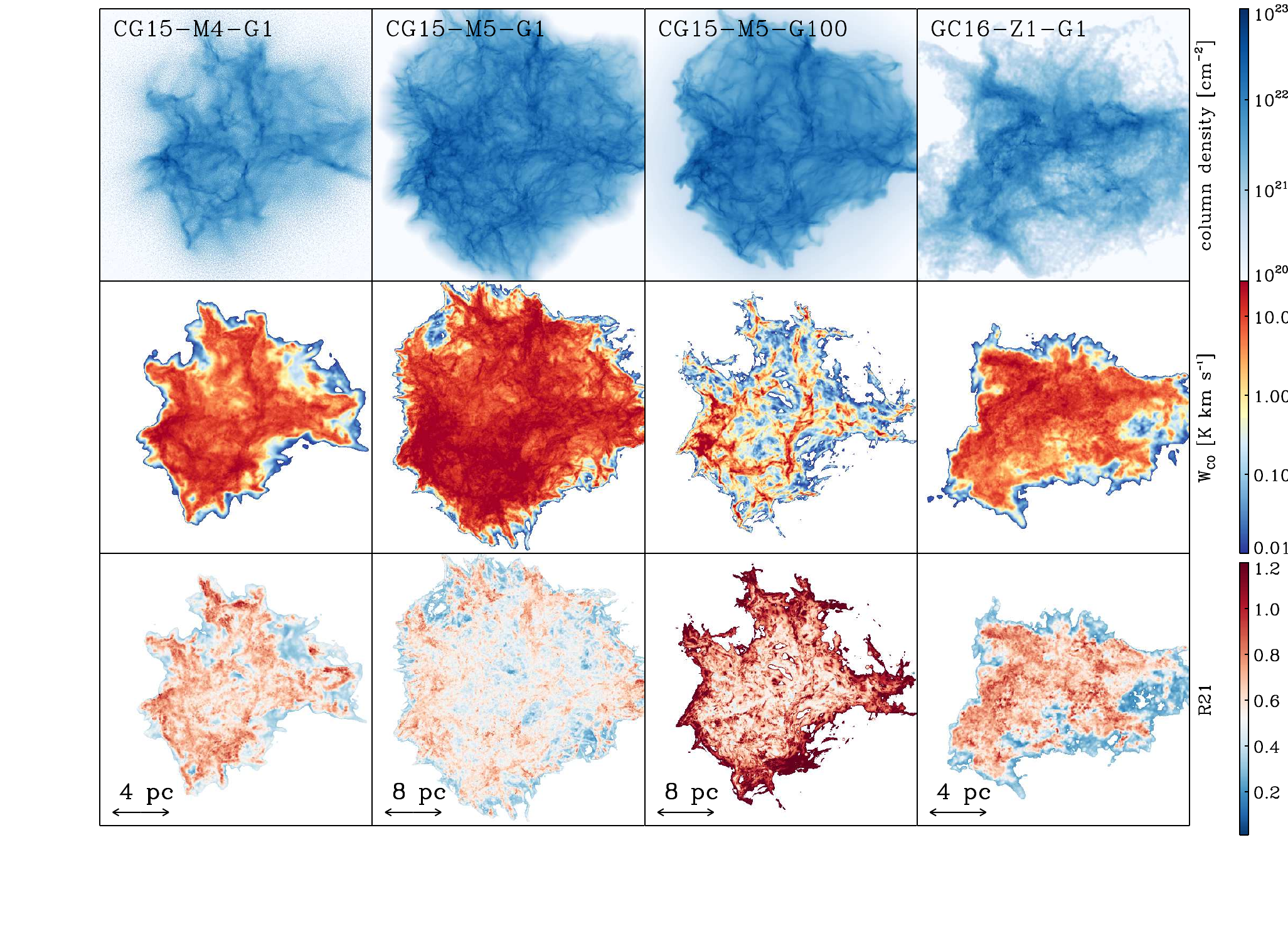}
\caption{Top row: Column density maps for simulations CG15-M4-G1, CG15-M5-G1, CG15-M5-G100 and GC16-Z1-G1. Middle row: The integrated intensity of the CO $J=2-1$ transition for each simulation. Bottom row: The ratio, $R_{21}$, of the integrated intensity of the CO $J=2-1$ and $J=1-0$ transitions for each simulation.}
\label{fig:MAPS}
\end{figure*}

To qualitatively illustrate how the cloud morphology changes, Figure \ref{fig:MAPS} contains the following simulations CG15-M4-G1, CG15-M5-G1, CG15-M5-G100 and GC16-Z1-G1. In the upper panels of Figure \ref{fig:MAPS} we present the column densities at the time when the synthetic observations are created. The fact that these simulations all used the same random seed in the turbulent initial velocity field is evident in the column densities images. The middle panels show the synthetic observations for the $^{12}$CO ($J=2-1$) line. The synthetic observations are able to recover the general structure of the cloud, however, the filamentary structures seen in the column density maps are not as easily identified in the emission maps, due to the optically-thick nature of the CO emission lines. 

Comparing the first two columns of Figure \ref{fig:MAPS} reveals how changing the mass of the cloud alters their emission. By increasing the mass by a factor of 10, but maintaining the initial density, the size of the cloud is effectively doubled. Since the initial virial state of the simulations is the same, the initial velocity dispersion is roughly 5 times higher in CG15-M5-G1 than in CG15-M4-G1. As the sound speed of the gas is similar in the two simulations, the higher-mass cloud has Mach number 4-5 times higher than the low mass cloud.  As a consequence, the higher-mass cloud has a more intricate web-like structure with a larger number of turbulence driven filaments. This is reflected in the synthetic images of the emission, since high density regions correlate with high intensity regions and vice-versa, and as we shall see this is an important factor in the distribution of $R_{21}$. 

The middle two columns compare the effect of varying the ISFR and the CRIR. In the high ISRF/CRIR scenario most of low-density, poorly-shielded CO has been dissociated and the thin, low density filaments have completely disappeared. More evident in the synthetic observations is how the apparent size of the cloud has been reduced by removing the low intensity regions that were previously enveloping the entire cloud. 

Finally the bottom panels of Figure \ref{fig:MAPS} contain the $R_{21}$ maps. As shown by \citet{2017MNRAS.465.2277P}, the area-weighted PDF of $R_{21}$ has a bimodal distribution, with peaks centred at $R_{21} \sim 0.3$ and $R_{21}\sim 0.7$. In the first two columns, it is clear that the ratio map is mostly dominated by values of $R_{21}\sim 0.7$, nonetheless lower values of $R_{21}$ are present in regions where $ W_{21} < {\rm few\ K\ km\ s^{-1}}$. A very different picture is seen in the $R_{21}$ map for CG15-M5-G100:  towards the centre of the cloud $R_{21}\sim 0.5 - 0.7$ but at the outskirts of the cloud $R_{21} > 1$. The high ISRF results in very high temperatures ($ T > 40 {\rm K}$) at the edge of very dense ($n > 10^3 {\rm cm ^{-3}}$) regions of the cloud. In such circumstances, the $\tau = 1$ surface for $J=1-0$ can be deeper into the cloud than the $\tau = 1$ surface for $J=2-1$, which can result in $R_{21} > 1$. This effect was demonstrated in the 1-D models of  \citet{1994ApJ...425..641S}. It is worth noting that $R_{21}$ can also be larger than 1 when the source of radiation is embedded within the cloud (see Figure 11 from \citealt{2015ApJS..216...18N}). Finally when comparing $R_{21}$ between CG15-M4-G1 and GC16-Z1-G1 it is clear that the morphology of the cloud (and thus the choice of turbulent velocity field) has an impact on the final value of $R_{21}$.

Qualitatively the ISRF and the CRIR have the biggest impact on $R_{21}$'s value and distribution. Their combined effects can hinder the accuracy of adopting a constant value of $R_{21}\sim 0.7$ and therefore over or underestimate the derived value for $W_{10}$.

\subsection{Systematic dependences of $R_{21}$}

\begin{figure*}
  \centering
  \subfloat{\includegraphics[width=0.5\linewidth]{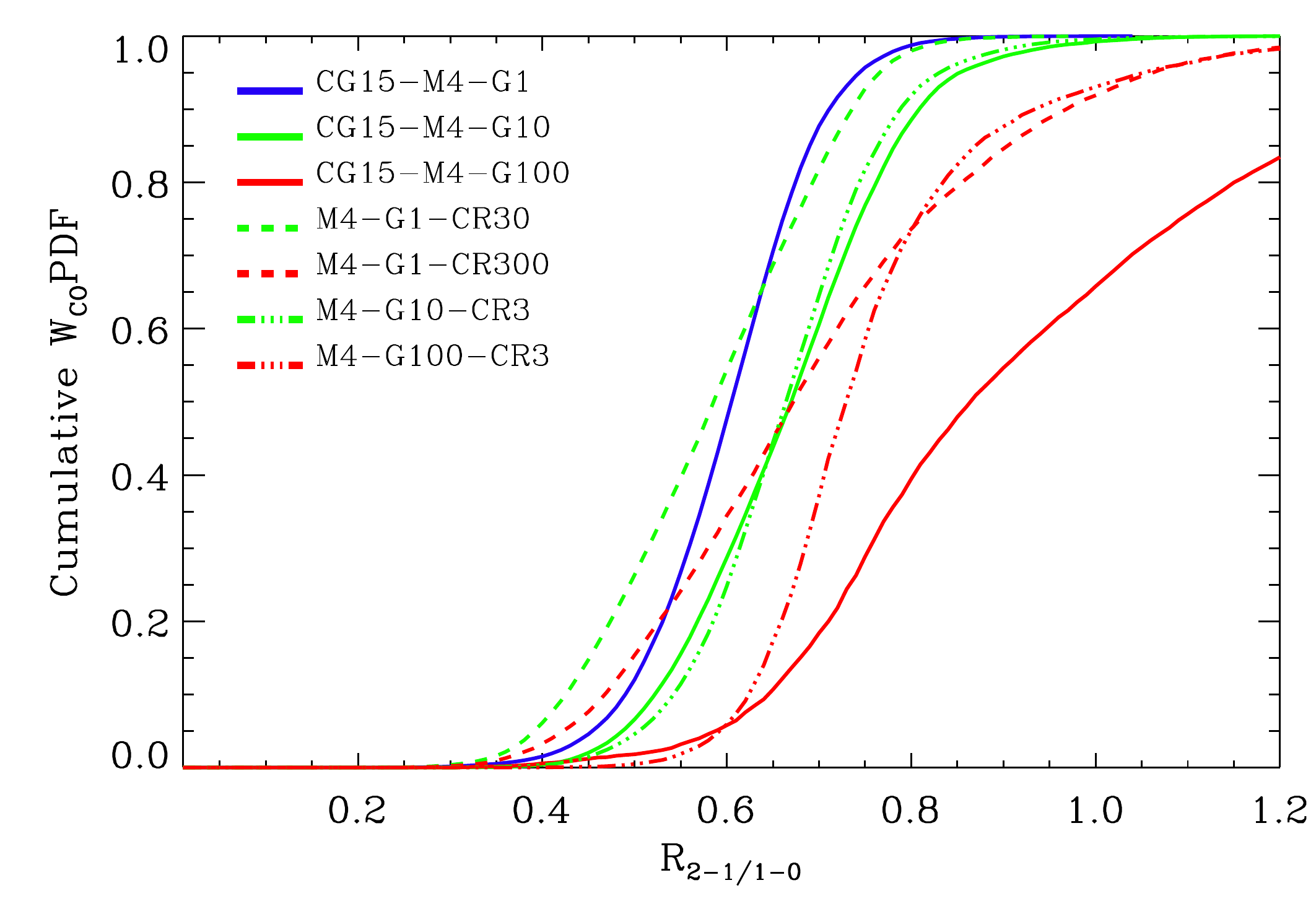}}
  \hfill
  \subfloat{\includegraphics[width=0.5\linewidth]{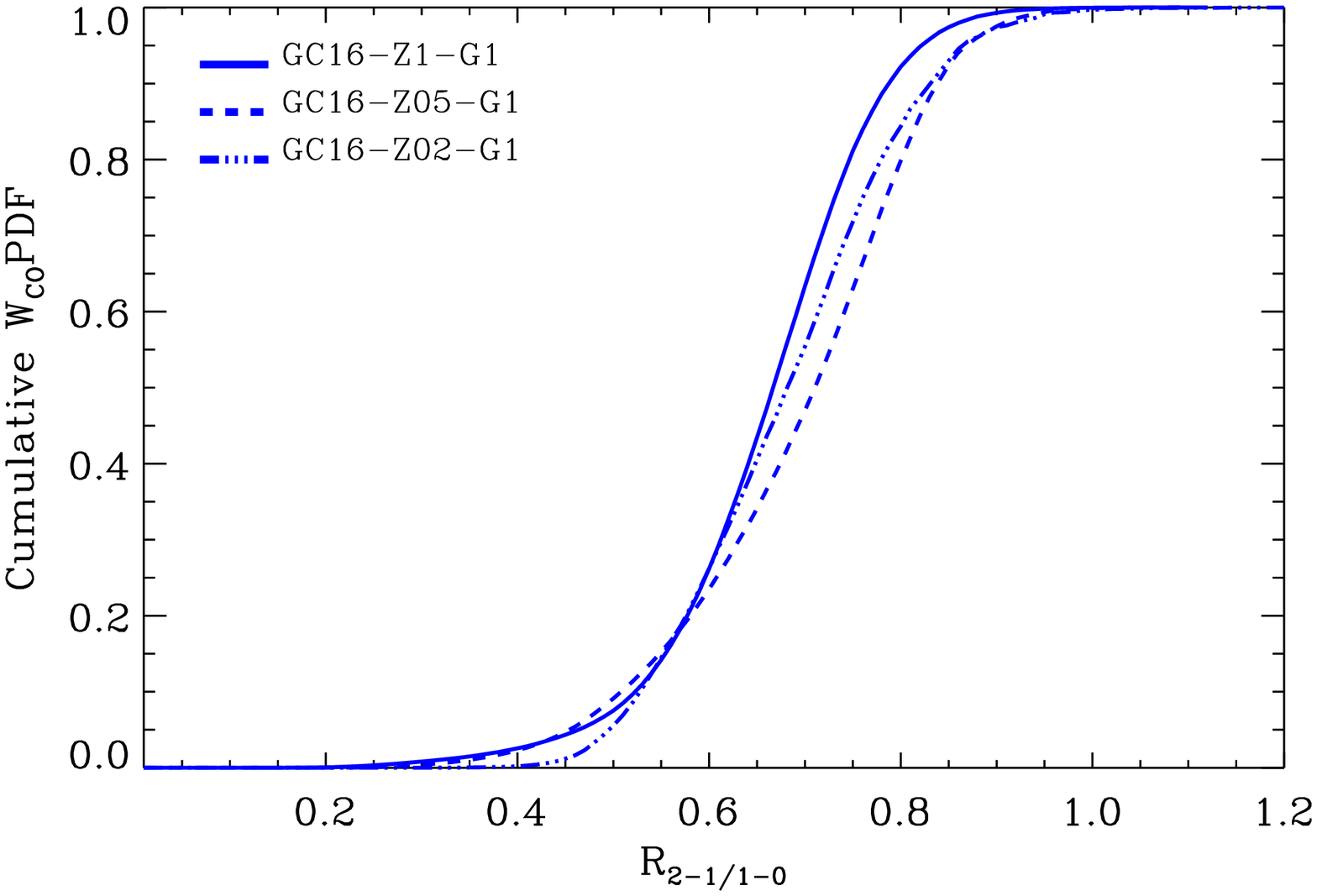}}
  \hfill
  \subfloat{\includegraphics[width=0.5\linewidth]{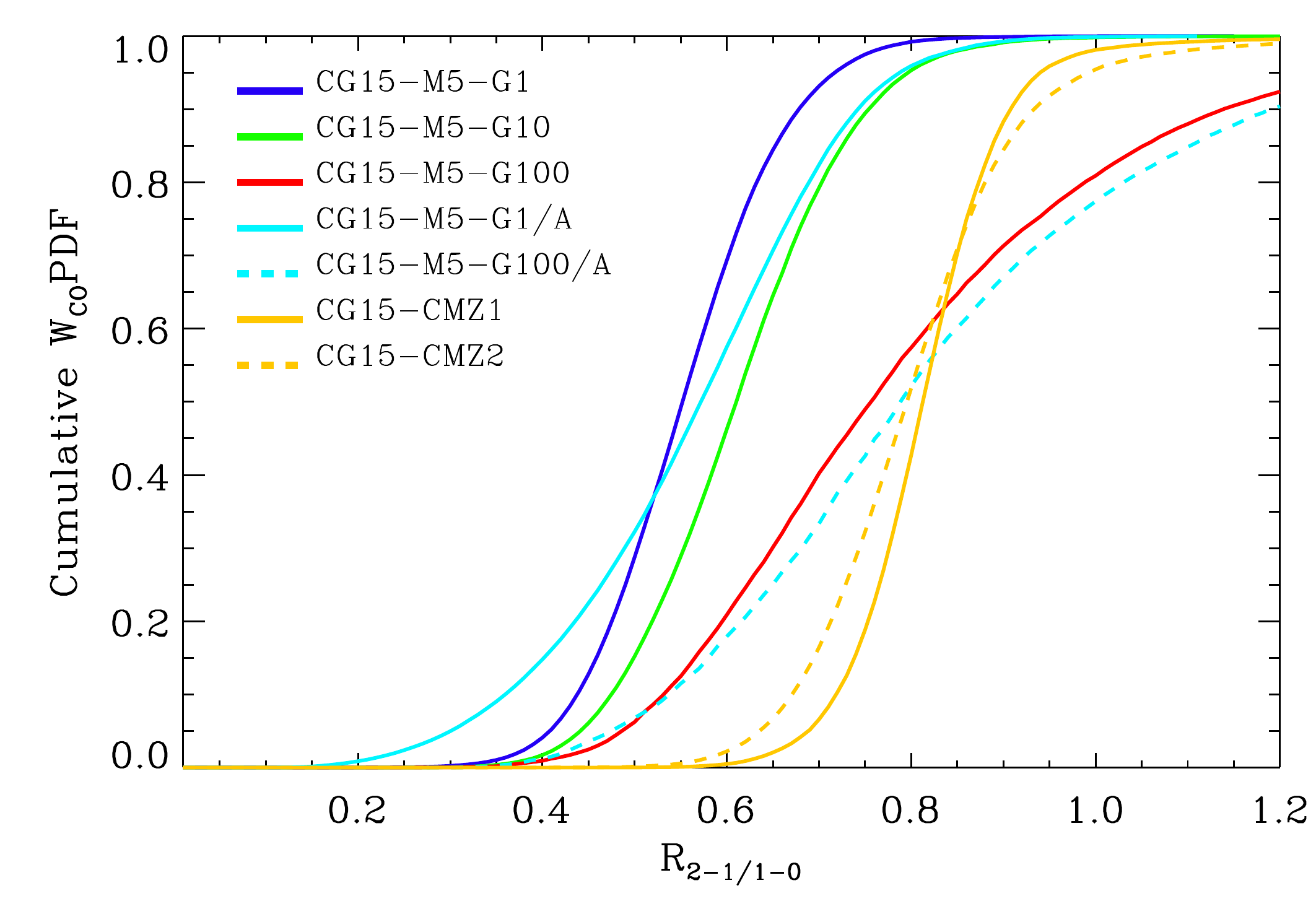}}
   \hfill
  \subfloat{\includegraphics[width=0.5\linewidth]{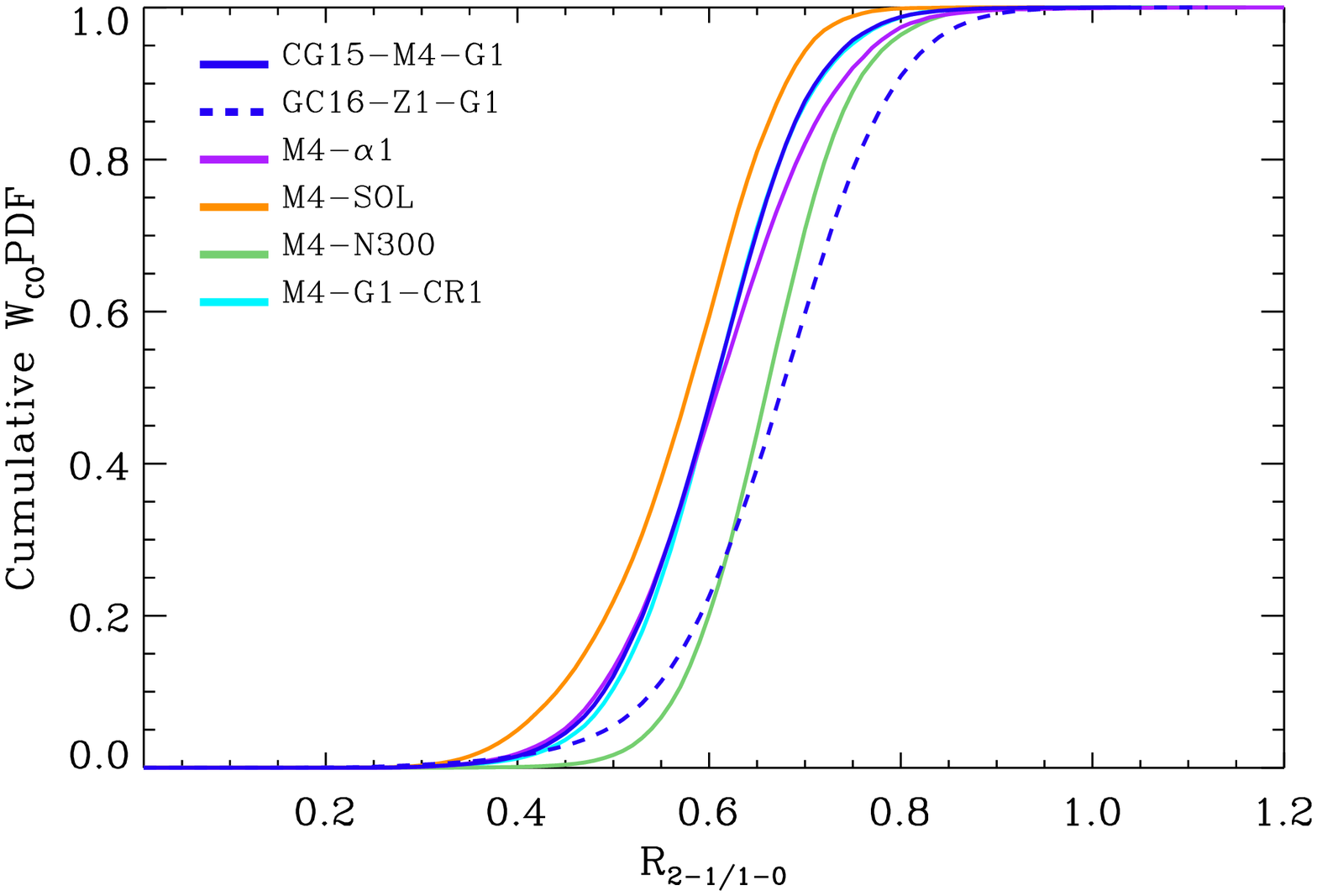}}
  \caption{ The cumulative PDF of $R_{21}$ weighted by W$_{\rm CO}$ for different sets of clouds grouped by variations in their physical parameters. Top left: Small clouds ($10^4 M_{\odot}$), at solar metallicities and with varying ISRF, CRIR or both. Top righ panel: Small clouds ($10^4 M_{\odot}$), with a solenoidal turbulent seed and varying metallicities. Bottom left panel: Large clouds ($10^5 M_{\odot}$) with variations in both ISFR and CRIR.  Cyan lines are clouds that start atomic and yellow lines clouds with initial $n=10^4\ {\rm cm^{-3}}$ Bottom right panel: Small clouds ($10^4 M_{\odot}$) with changes to $\alpha_{vir}$, initial density or the turbulent seed }
  \label{fig:WPDFvsR21}
\end{figure*}

The morphology of the cloud gives a qualitative picture of the impact that initial mass, turbulence and the ISRF/CRIR have on the CO emission and the corresponding line ratios. In this section we focus on the $R_{21}$  line ratio. The different effects on the line ratio can be quantified by plotting the cumulative PDF of $R_{21}$ weighted by integrated intensity (Figure \ref{fig:WPDFvsR21}). The separation between the curves for each cloud shows how changes in environmental conditions impact the distribution and average value of $R_{21}$. We reiterate here that low values of $R_{21}$ ($\sim 0.2-0.4$) are associated with warm and diffuse gas, while high values of the $R_{21}$ ($\sim 0.6-0.8$) are associated with cold and dense gas \citep{2017MNRAS.465.2277P}.  

\subsubsection{The ISFR and the CRIR}

The top left panel of Figure \ref{fig:WPDFvsR21} plots the cumulative PDF for a set of the low mass clouds at solar metallicities. As the ISFR and the CRIR increase (solid lines), a larger fraction of the overall emission is associated with larger values of $R_{21}$. In this case almost half of the emission is associated with line ratio values of $R_{21} > 1$. As the ISRF increases the unshielded molecular gas is fully dissociated, destroying most of the CO in the already diffuse gas and therefore resulting in no CO emission from these regions of the cloud. This is consistent with the fact that lower values of $R_{21}$ ($\sim 0.3$) are associated with diffuse and warm areas of the cloud \citep{1994ApJ...425..641S,2017MNRAS.465.2277P}. Similar to CG15-M5-G100 (see bottom left panel and Figure \ref{fig:MAPS}), values of $R_{21} > 1$ are correlated with CO emission originating from dense and hot gas at the edge of the cloud. 

The dashed-dotted lines represent the runs where only the ISRF field was increased while the CRIR was left unchanged. For these clouds a large fraction of the emission is also associated with higher line ratios. However, a larger fraction (about 80 \%) of the overall emission is associated with line ratios of $R_{21} \sim 0.6-0.8$. A smaller CRIR means the dense gas within the shielded regions of the cloud is not being heated and therefore results in a more compact cloud. As such most of the emission originates from dense and cold gas, which explains why a larger fraction of the emission is correlated with $R_{21} \sim 0.7$.

On the other hand the dashed lines represent the clouds where the ISRF was left constant while the CRIR were increased.  In this case about  50 \%  of the overall emission is correlated with lower line ratios ($R_{21} \sim 0.3-0.6$). As shown by \citet{2016MNRAS.456.3596G} (see also \citealt{2015ApJ...803...37B,2017ApJ...839...90B}), an increase in $\zeta_H$ can lead to a decrease in CO abundance. Even though the total abundance of CO has been reduced it is still well shielded from the UV rays, resulting in emission originating from low density gas and therefore associated with lower line ratio values. The effect of $\zeta_H$ on CO abundance will be more thoroughly discussed in Section \ref{dissec}. Additionally, the high CRIR will heat up dense regions of the cloud, thereby increasing the overall emission associated with higher line ratios ($R_{21} > 0.9$).

Finally, by comparing the bottom left panel  with the top left panel, we can see that the compound effect of the ISRF and the CRIR does not depend on the cloud mass. Although the higher mass clouds tend to contain more filamentary structures (as discussed in the previous Section), the combination of high ISRF and high CRIR results in heating which acts to smear out these high-density structures in both cases. As such, the variation in $R_{21}$ as function of ISRF + CRIR looks very similar in the high- and low-mass clouds.

\subsubsection{Metallicity}

The top right panel of Figure \ref{fig:WPDFvsR21} plots the cumulative PDF of low-mass clouds ($10^4 M_{\odot}$), with a solenoidal turbulent seed and varying metallicities. By reducing the metallicity, the conditions to form CO tend to only be achieved within dense cores  \citep{2016MNRAS.456.3596G}, which results in less CO emission from diffuse regions of the cloud. This can reduce the percentage of the low values of $R_{21}$ that are associated with low density gas, and indeed we see that in the case of $Z = 0.2\,Z_{\odot}$, there is almost no emission associated with $R_{21} < 0.4$. 

We also see variations in the fractions of emission associated with the higher $R_{21}$ values, although the overall effect is still quite small.  These variations are likely due to the fact that as we decrease the metallicity, the abundance of both CO and H$_2$ (the dominant collisional partner for CO) are both decreasing. As such, for a given total density of gas, the {\em effective} density of the species contributing to the emission is falling as we decrease the metallicity, which has the result of reducing $R_{21}$. However, this is going to be partly offset by the fact that the gas is hotter at lower metallicities, which acts to raise $R_{21}$, potentially explaining the overall small variation in the cumulative distribution of $R_{21}$. 

\subsubsection{Mass and molecular fraction}

The bottom left panel of Figure \ref{fig:WPDFvsR21} contains all the clouds with total mass of $10^5 M_{\odot}$. The first point to notice is, by comparing the blue lines in the top and bottom left panels, that increasing the total mass of the cloud leaves the overall value and distribution of $R_{21}$ relatively unchanged. 

The effect of having an initially atomic cloud is on the fraction of the overall gas associated with lower line ratios, where about 40 \% of the emission has values of  $R_{21} < 0.5$ for clouds that start fully atomic. This tail at low values of $R_{21}$ is correlated with a larger fraction of the emission coming from low density/high temperature regions. This is consistent with the idea that less molecular material at the beginning will delay the formation of CO \citep{Glover:2012dd}, resulting in lower column densities of CO. Given that the impact of the ISRF and the CRIR are much stronger, as discussed above, the effect of initial $f({\rm H_2})$ can be more easily observed at low ISRF, as the high-ISRF cases (solid red and dashed cyan) are relatively similar.

Finally the two CMZ-like runs with an initial density of $n=10^4\ {\rm cm}^{-3}$ have a larger average value of $R_{21}$, where most of the overall emission is associated with $R_{21} \sim 0.8$. This is due to the fact that the initial density is well above the critical density of the first two rotational transition lines of CO ($n_{\rm crit,1-0} \approx 2000 {\rm cm}^{-3} $, $n_{\rm crit,2-1} \approx 10^4 {\rm cm}^{-3} $). As a result the gas is well shielded from the ISRF and CO is easily excited. Since most of the emission is originating from high density gas, this means that $R_{21}$ will be centred near $\sim 0.7$. 

\subsubsection{Turbulence, $\alpha_{\rm vir}$ and initial density}

As mentioned before we have used clouds from \citet{2015MNRAS.452.2057C} and \citet{2016MNRAS.456.3596G}, however as it can be seen in Table \ref{tab:InitialConditions} that these have slightly different initial conditions and environmental parameters. The bottom right panel of Figure \ref{fig:WPDFvsR21}  hows the clouds that explore variations in only these features. First we must note that even though CG15-M4-G1(blue solid line) and GC16-Z1-G1(blue dashed line) have very similar initial conditions, the changes in turbulence, $\alpha_{\rm vir}$ and initial density still noticeably change $R_{21}$'s distribution. 

\citet{2016MNRAS.456.3596G} adopted a solenoidal velocity field. Such a field field has no compressive motions ($ \nabla \cdot {\bf u} = 0 $), resulting in more flocculent cloud (lower density) with a narrower PDF than is found for fields with similar Mach numbers and power spectra, but which contain compressive modes \citep{2010A&A...512A..81F}. Consequently, a larger fraction of the emission originates from more diffuse gas, which is correlated with lower values of $R_{21}$ in the clouds which adopt a solenoidal velocity field.

Increasing the value of $\alpha_{\rm vir}$ slightly increases the average value of $R_{21}$. A larger kinetic energy results in higher Mach numbers, and thus more shock-compressed high-density regions. Although the higher $\alpha_{\rm vir}$ will also cause the clouds to expand, which would lower  $R_{21}$, there is not sufficient time for this to occur in our study; these clouds form stars more rapidly than the lower $\alpha_{\rm vir}$ clouds, which means we post-process them at an earlier time, before they have had the chance to expand.  The net effect is more gas is at high densities when we come to post-process, and thus a slightly larger fraction of the cloud will be associated with larger values of $R_{21}$.  Clearly this effect will depend on time, which is beyond the scope of this study. 

The green line shows, to a lesser extent, what was observed for CMZ-like clouds; increasing the initial density reduces the amount of diffuse gas and effectively increases the fraction of the gas associated with higher values of $R_{21}$. The slightly different $\zeta_{\rm H}$ between CG15-M4-G1 and GC16-Z1-G1 has very little effect, this can be seen by the cyan line where $\zeta_{\rm H} = 1 \times 10^{-17} {\rm s^{-1}}$. Given that the variation in $\zeta_{\rm H}$ is small the change in $R_{21}$ is not substantial. However reducing the CRIR does slightly reduce the amount of emission associated with low line ratios. Finally, we note that the combined effect of turbulence, $\alpha_{\rm vir}$, initial density and slightly different $\zeta_{\rm H}$ explain the different average value and distribution of $R_{21}$ between CG15-M4-G1 and GC16-Z1-G1.

\section{$R_{21}$ from observationally unresolved clouds}\label{ressec2}

The previous section explained how variations in the initial conditions and in the surrounding environment affect both the mean value and distribution of $R_{21}$. Moreover, effects on $R_{21}$ can be associated with different regions of the cloud and are correlated with local physical properties of the gas. However, when comparing to observations in an extragalactic context, where $R_{21}$ is used as a conversion factor, details about the varying value of $R_{21}$ within molecular clouds are not known. In this context one of our clouds will most likely be smaller than the observational beam size (c.f. \citealt{ 2016ApJ...822L..26B,2013ApJ...779...42S}) and therefore the total intensity of a one or more GMCs will be averaged within this beam. Therefore to study $R_{21}$ in the context of observationally unresolved clouds, we first take the area-weighted intensity average of each cloud for each rotational transition line and then take the ratio of the two averaged intensities.  In this section we focus on presenting the results from these area-averaged line ratios.

\subsection{Averaged $R_{21}$ for the whole cloud}

\begin{figure}
\centering
\includegraphics[width=1.\linewidth]{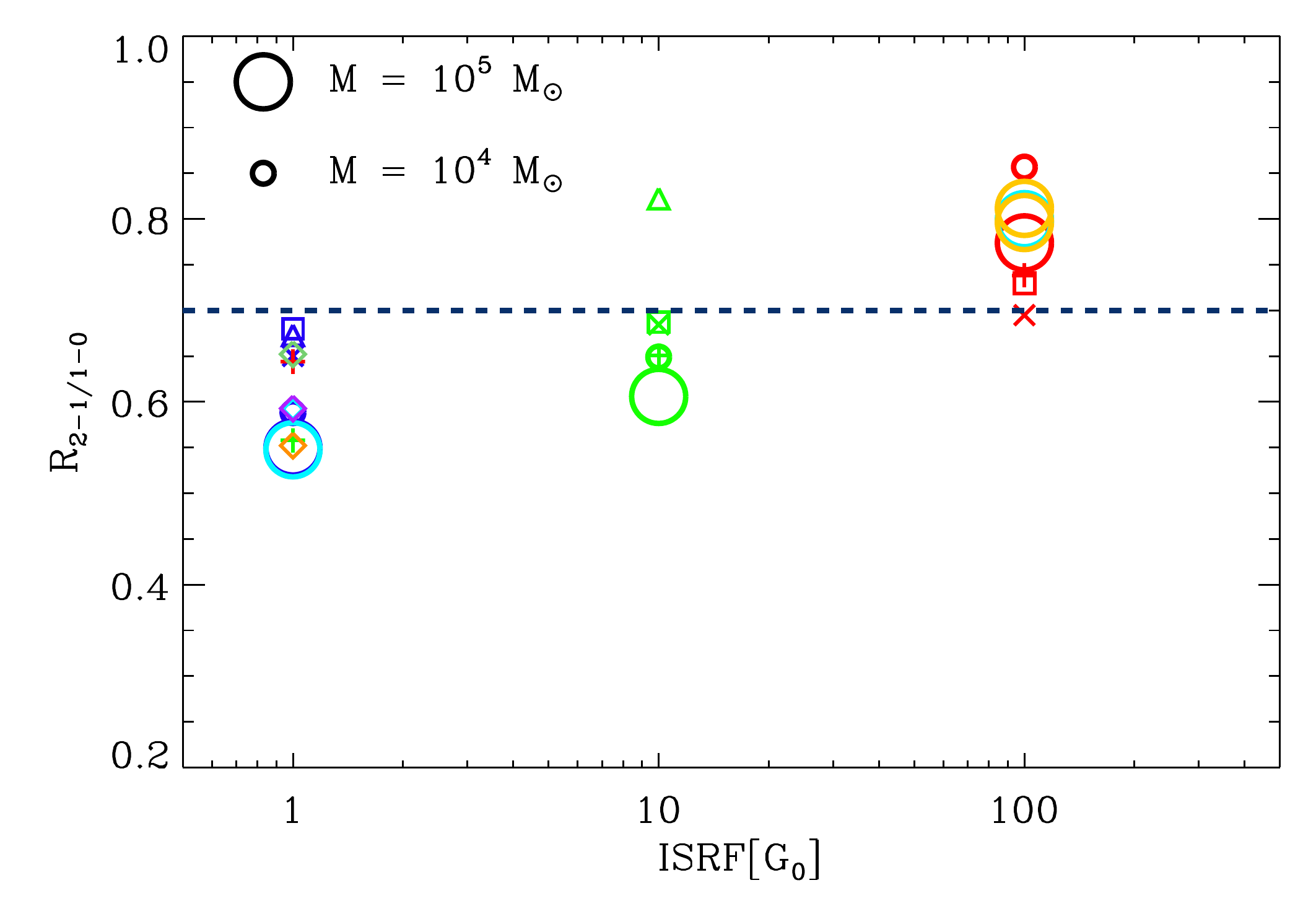}
\caption{The averaged value of $R_{21}$ for each cloud as a function of their respective ISRF ($G_0$). Large circles represent clouds that have a mass of $M = 10^5\ M_{\odot}$ and small shapes represent clouds with a mass of $M = 10^4\ M_{\odot}$.  Blue, green and red shapes represent an increase of the ISRF and/or CRIR by 1,10 or 100 respectively. Xs, squares and triangles represent metallicities of $Z_{\odot}$, $0.5 Z_{\odot}$ and $0.2 Z_{\odot}$ respectively. Large cyan circles denote an initial hydrogen fraction of $f({\rm H_2})=0$. Yellow circles are the two runs with an initial number density of  $n=10^4\  {\rm cm}^{-3}$. Plus signs are clouds where the ISRF and the CRIR have been varied independently. Diamonds are the additional runs plotted in the bottom right panel of Figure \ref{fig:WPDFvsR21} and have the same colours. Finally the dashed line represents the standard value used for converting $W_{21}$ to $W_{10} $. Note: we have not included GC16-Z02-G100 since the gas is not able to form enough molecular gas and therefore produces almost no CO emission} 
\label{fig:R21vsSFR}
\end{figure}

In the previous section it was suggested that varying the UV intensity has an impact on $R_{21}$. We therefore plot the area-averaged value of $R_{21}$ against the ISRF for all the clouds in Figure \ref{fig:R21vsSFR}. 

First thing to note is that the averaged value of $R_{21}$ covers a range of values between $0.5 - 0.9$, this confirms that changes in the cloud's environmental conditions do influence the overall value of $R_{21}$. An increasing ISRF is directly correlated with an increase in the average value of $R_{21}$. Additionally, within each ISRF bin there is appreciable scatter, which can be attributed to the different changes in environmental conditions discussed in the previous Section. However, since the plot shows an averaged value for $R_{21}$, the specific changes in the distribution of $R_{21}$ are reduced making it harder to distinguish between different environmental effects. 

Extragalactic observations often use $R_{21}$ as a conversion factor rather than a diagnostic tool for GMC structure. In that context the averaged dashed line in Figure \ref{fig:R21vsSFR} represents the observationally derived and most commonly used value of $R_{21}$ \citep{1990ApJ...348..434E,1991A&A...251....1C,1995A&A...303..851B,1997ApJ...486..276S,1997IAUS..170...39H,2001ApJS..136..189S,2008AJ....136.2846B,2009AJ....137.4670L,2011MNRAS.416.1250B}. This line lies in the middle of the scatter of $R_{21}$ values of our clouds, suggesting that $R_{21}\sim 0.7$ is a good first approximation for converting $W_{21}$ into $W_{10}$. Nonetheless, it questions the reliability and robustness of the conversion factor,  and suggests that adopting a fiducial value can lead to errors in derived quantities such as the total molecular gas. We discuss the consequences of such an approach, and possible solutions, in Section \ref{dissec}.

It is important to note that the agreement between our simulations and the accepted value of $R_{21}$ improves when considering higher sensitivity cuts to our detection limits. Our synthetic observations have an already low emission cut of $0.01\ {\rm K\ km\ s} ^{-1}$, when we impose a detection limit of $5\ {\rm K\ km\ s}^{-1}$ instead the scatter is almost completely gone. Considering that the emission from diffuse gas associated with lower line ratios is always very faint, it follows that $R_{21} \sim 0.7$ as sensitivity is reduced. This is a consequence of $R_{21}$ being derived in a galactic context where clouds are well resolved, and suggests that sensitivity plays an important role in how we have empirically derived the accepted value of  $R_{21}$.

\subsection{CO emission as a probe of physical conditions}

Naturally CO emission is directly correlated to the temperature of CO molecular gas within the cloud. However, the overall temperature range probed by CO is not necessarily representative of the range of temperatures found in  ${\rm H_2}$-rich gas. In the section, we explore whether CO emission, and more explicitly $R_{21}$, can accurately trace the gas temperature within GMCs. In order to compare to the averaged value of $R_{21}$ calculated before, we define the cloud-averaged temperature as 

\begin{equation}
\langle T\rangle =  \frac{\sum\limits_{i=1}^{n} m_i T_i \chi_i }{\sum\limits_{i=1}^{n} m_i \chi_i}
\label{eqn:TH2}
\end{equation}

\noindent where the sum is over all the SPH particles, $m$ is the particle mass, $T$ is the particle temperature and $\chi$ is the abundance of the molecule of interest. In this case we calculate the average temperature of ${\rm H_2}$ ($T_{\rm H_2}$) and of CO ($T_{\rm CO}$). It is important to note that the $T$ is the kinetic temperature of the gas obtained from the original SPH data that was the input for the for the radiative transfer.
 
\begin{figure}
  \centering
  \includegraphics[width=1.\linewidth]{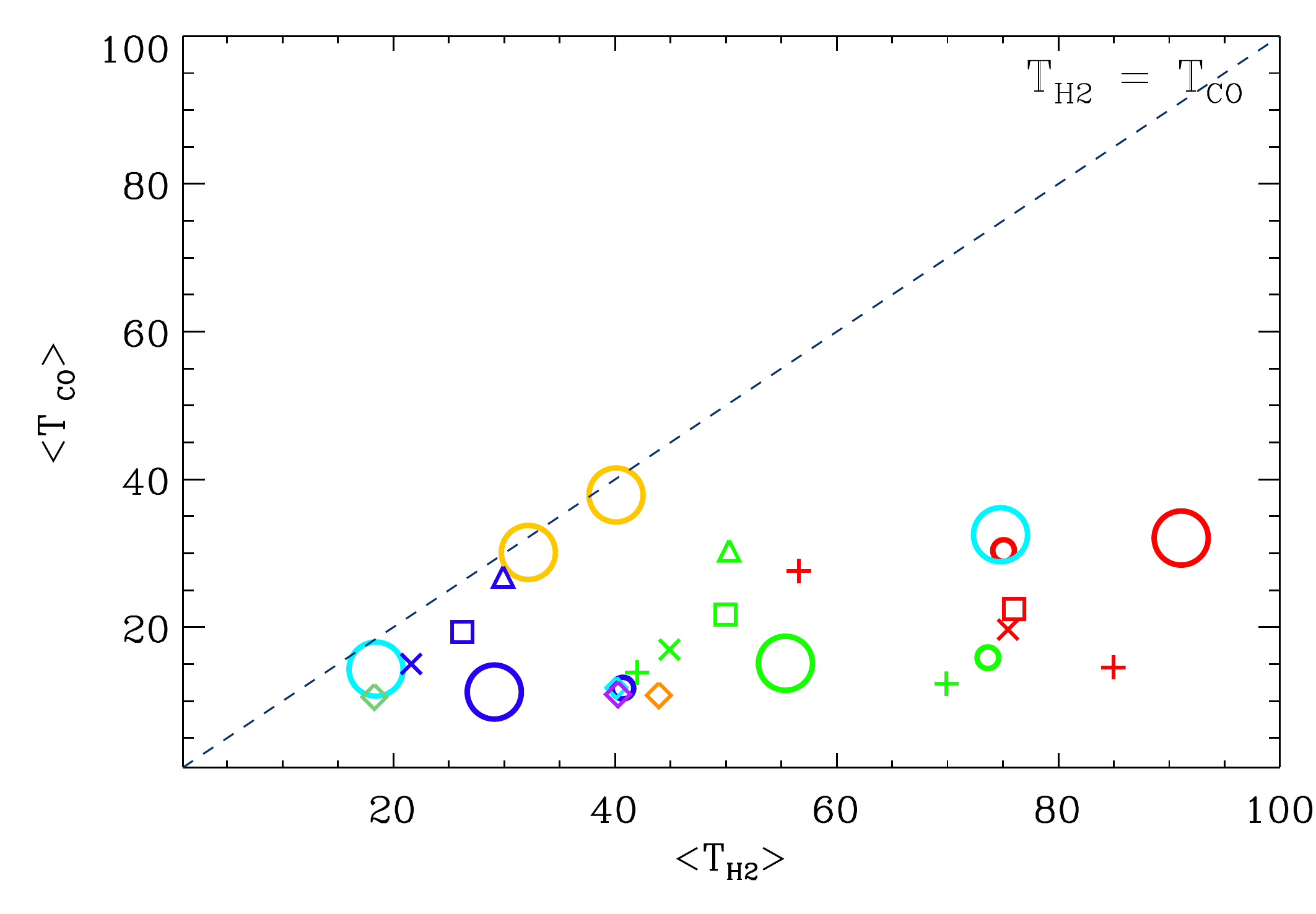}
  \caption{The average temperature of ${\rm H_2}$ as calculated by equation \ref{eqn:TH2} plotted against the average temperature of CO. Each point represents a cloud and has the same colour and shapes as in Figure \ref{fig:R21vsSFR} }
  \label{fig:R21vsTs}
\end{figure}

In Figure \ref{fig:R21vsTs} we plot $\langle T_{\rm H_2}\rangle$ vs $\langle T_{\rm CO}\rangle$ where every point represents a cloud and the points have the same shape and colour as in Figure \ref{fig:R21vsSFR}. First thing to note is that the temperature of the molecular cloud -- as defined by the H$_2$ -- increases as we increase the ISRF and/or CRIR, which is to be expected. However this increase in temperature is only reflected in $\langle T_{\rm H_2}\rangle$, while CO-rich gas never reaches temperatures above $T\sim 40$ K. As such the CO emission from the cloud is only tracing the temperature variations in a sub-set of the full H$_2$ molecular gas. This is easily understood since the bulk of the CO gas is within well-shielded regions where temperature of the gas is low and the densities are above $n_{\rm crit}$. This is not the case for H$_2$, which is present in more diffuse regions where the temperature range can be higher (see also \citealt{2015MNRAS.454..238W} and \citealt{2016MNRAS.462.3011G} who find similar results in larger-scale simulations of the ISM).

One notable exception is when considering the two CMZ clouds (yellow circles), in this case $T_{\rm CO}$ and $T_{\rm H_2}$ are very similar. Since the initial density of these clouds is much higher, this results in most of the gas being well shielded and therefore at similar temperatures. Since the densities are high enough to excite CO then the CO emission is well correlated with the overall temperature of the gas. As was mentioned in the previous Section this is also reflected in the small variations of $R_{21}$.

\subsection{Alternative line ratios}

Having created synthetic observations for the lowest three emission lines of CO's rotational ladder, it is a simple task to consider other ratios between these lines as possible conversion factors. In this section we explore the the ratios  $R_{32}$ and $R_{31}$, which have both been employed as conversion factors in extragalactic studies. 

\subsubsection{$R_{32}$}

\begin{figure}
\centering
\includegraphics[width=1.\linewidth]{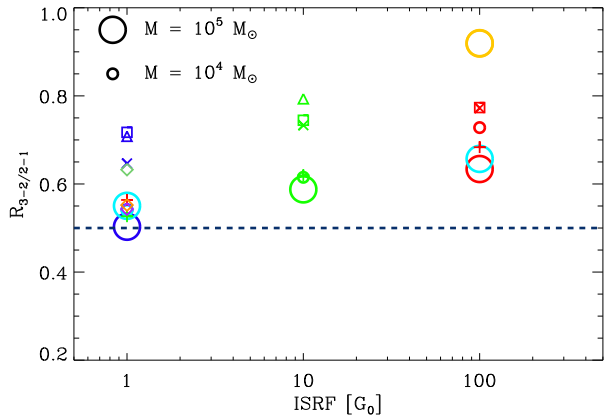}
\caption{Similar to Figure \ref{fig:R21vsSFR} but for the ratio of the $J=3-2$ and $J=2-1$ rotational transition lines. Colour coding is the same as in Figure \ref{fig:R21vsSFR}. The dashed line is the commonly used value for $R_{32}$ \citep{2013MNRAS.433.1837V}} 
\label{fig:R32vsSFR}
\end{figure}

We first consider $R_{32}$, that is the ratio between the third ($J=3-2$) and second ($J=2-1$) rotational levels of CO. In order to judge how $R_{32}$ varies we plot a similar figure to Figure \ref{fig:R21vsSFR}, where we take the average intensities of each line for each cloud and then calculate the line-ratio. This is shown in Figure \ref{fig:R32vsSFR}.  

In this case the overall scatter is considerably larger than it was for $R_{21}$ and our results are all above the standard value of $R_{32} =0.5$ \citep{2013MNRAS.433.1837V}. At the same time $R_{32}$ is also highly dependent on the changes in the ISRF as demonstrated by the increase in the averaged value of $R_{32}$ with increasing ISRF.  When we examine the properties of the individual clouds within each ISRF bin, the spread seems to be correlated with the initial density of the simulations. Considering that the critical densities of $J=2-1$ and $J=3-2$ are of the same order of magnitude ($n_{\rm crit} \sim 10^4\ {\rm cm^{-3}}$) then a larger fraction of $R_{32}$'s distribution originates from regions that are sub-critically excited i.e. $n < 10^4\ {\rm cm^{-3}}$ with lower initial densities. This explains why the value of $R_{32}$ is larger for most of the GC16 clouds that have a slightly larger initial density. 

Finally the CMZ-like clouds have a larger value of $R_{32} \sim  0.9$. Since the cloud starts with an initial density of $n= 10^4 {\rm cm}^{-3}$, which is close to the critical densities for both transitions, this results in comparable emission rates from both lines and therefore a ratio closer to unity. 

\subsubsection{$R_{31}$}

\begin{figure}
\centering
\includegraphics[width=1.\linewidth]{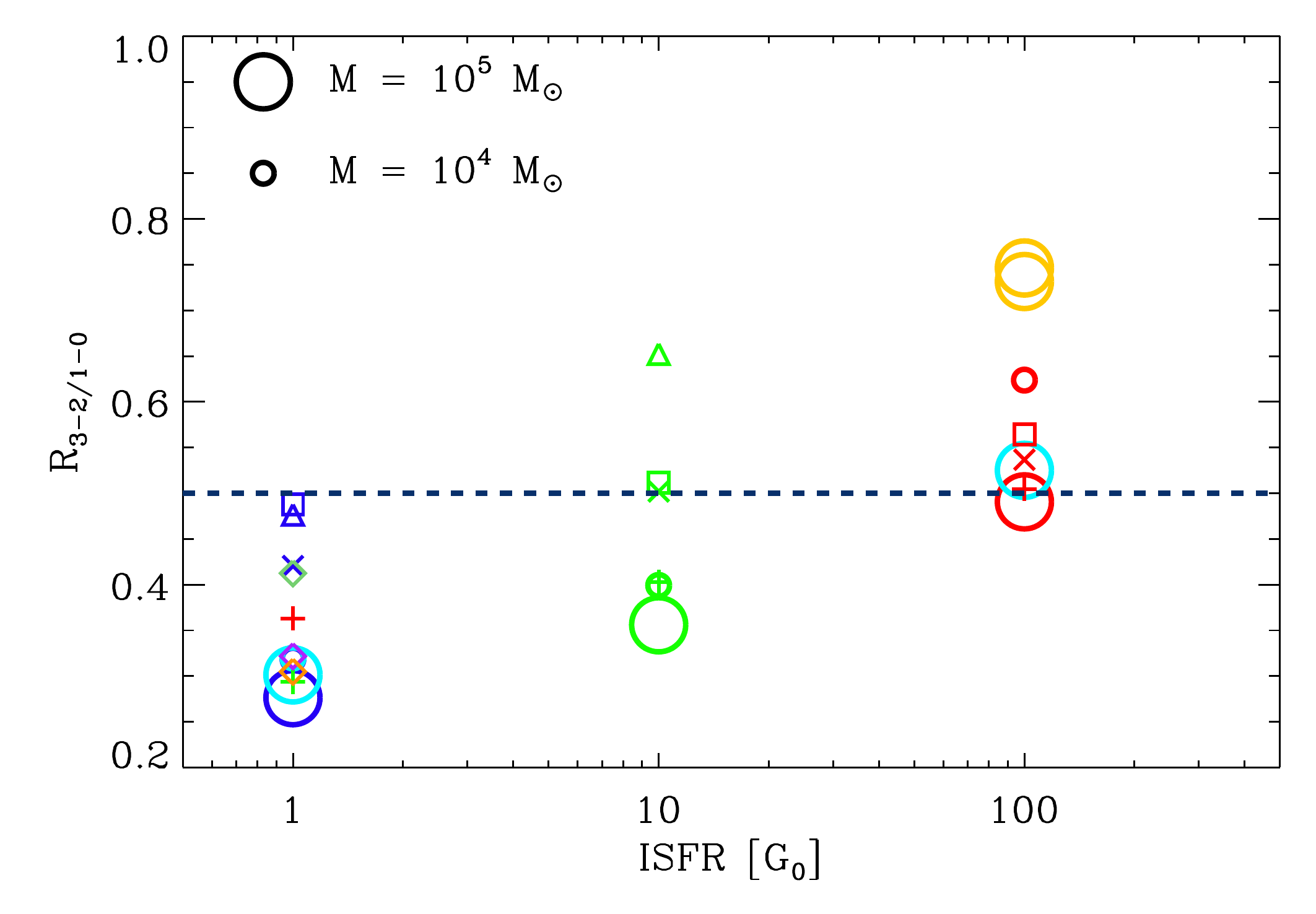}
\caption{Similar to the top panel of Figure \ref{fig:R21vsSFR} but for $J=3-2$ and $J=1-0$ rotational transition lines. The dashed line represents the commonly used conversion factor for $R_{31}$ \citep{2014MNRAS.442..558A}} 
\label{fig:R31vsSFR}
\end{figure}

In Figure \ref{fig:R31vsSFR} we look at the variations of $R_{31}$, that is the ratio between CO's third ($J=3-2$) and first ($J=1-0$) rotational emission lines. The scatter for $R_{31}$ is similar to $R_{32}$ and we also see an increase in the average value with increasing ISRF.  However, the value of $R_{31}$ is more evenly spread in each ISRF bin, suggesting that $R_{31}$ is slightly more susceptible to changes of environment and initial conditions of GMCs. 

The reason $R_{31}$ has a higher variability is due to the larger difference in excitation conditions for both lines. The critical density of the $J=3-2$ emission line is $n_{\rm crit} = 3.6 \times 10^4\ {\rm cm}^{-3}$ that is over an order of magnitude higher than  the $J=1-0$ line. Furthermore the difference in energy required to excite both lines is $\sim 27.7$ K, which is quite significant when considering the low temperature environments of dense regions within GMCs. This explains why at lower ISRF $R_{31}$ is small, since a significant amount of emission is arising from diffuse regions where the $J=1-0$ line is easily excited but $J=3-2$ is not. On the other hand at high ISRF both lines will be excited in dense regions, since the clouds are warmer, and in addition, the dissociation of the CO is the diffuse gas will  lower the contribution to the emission from $J=1-0$. Overall, the net effect is to increase the value of $R_{31}$. 

$R_{31}$ is usually taken to be $R_{31}= 0.5 \pm 0.2$. Our results show that the scatter of the average value for different clouds is just over the expected error for $R_{31}$. However, the results in Figure \ref{fig:R31vsSFR} also suggest that using $R_{31} =0.5$ can considerably underestimate or overestimate the amount of emission associated with $W_{10}$ , and in turn, any derived properties of the source. At the same time this suggests that if the value of the ISRF is known a better constrained conversion factor with a smaller error can be used. We shall explore this in the following Section.

\section{Discussion}\label{dissec}

\subsection{ $X_{\rm CO}$ in unresolved clouds }

The results presented in Section \ref{ressec2} show the dependence of the line ratios on the properties of the gas in the clouds, and the environmental conditions to which the clouds are exposed. In the Milky Way, where it is possible to resolve clouds, the line-ratios can serve as a probe of the physical conditions in the cloud. We have also shown that the area-averaged line-ratios arising from unresolved clouds also show significant variation. This can have implications for how they are used as conversion factors from the $J = 2-1$ and $J = 3-2$ lines to the more commonly used $J = 1-0$ transition that is used in the X-factor.  In this section, we explore how these variations in $R_{21}$ and $R_{31}$ affect our derivations of physical clouds properties. 

The total column density of ${\rm H_2}$ is often calculated by 

\begin{equation}
N_{\rm H_2} =  \frac{X_{\rm CO}W_{21}}{R_{21}}.
\label{eqn:xcor21}
\end{equation}

\noindent We therefore want to compare $N_{\rm obs}$ as calculated by Equation \ref{eqn:xcor21} with $N_{\rm real}$, where we take $N_{\rm real}$ to be the column density of ${\rm H_2}$ directly from the GADGET-2 snapshot. 

\begin{figure}
\centering
\subfloat{\includegraphics[width=1.\linewidth]{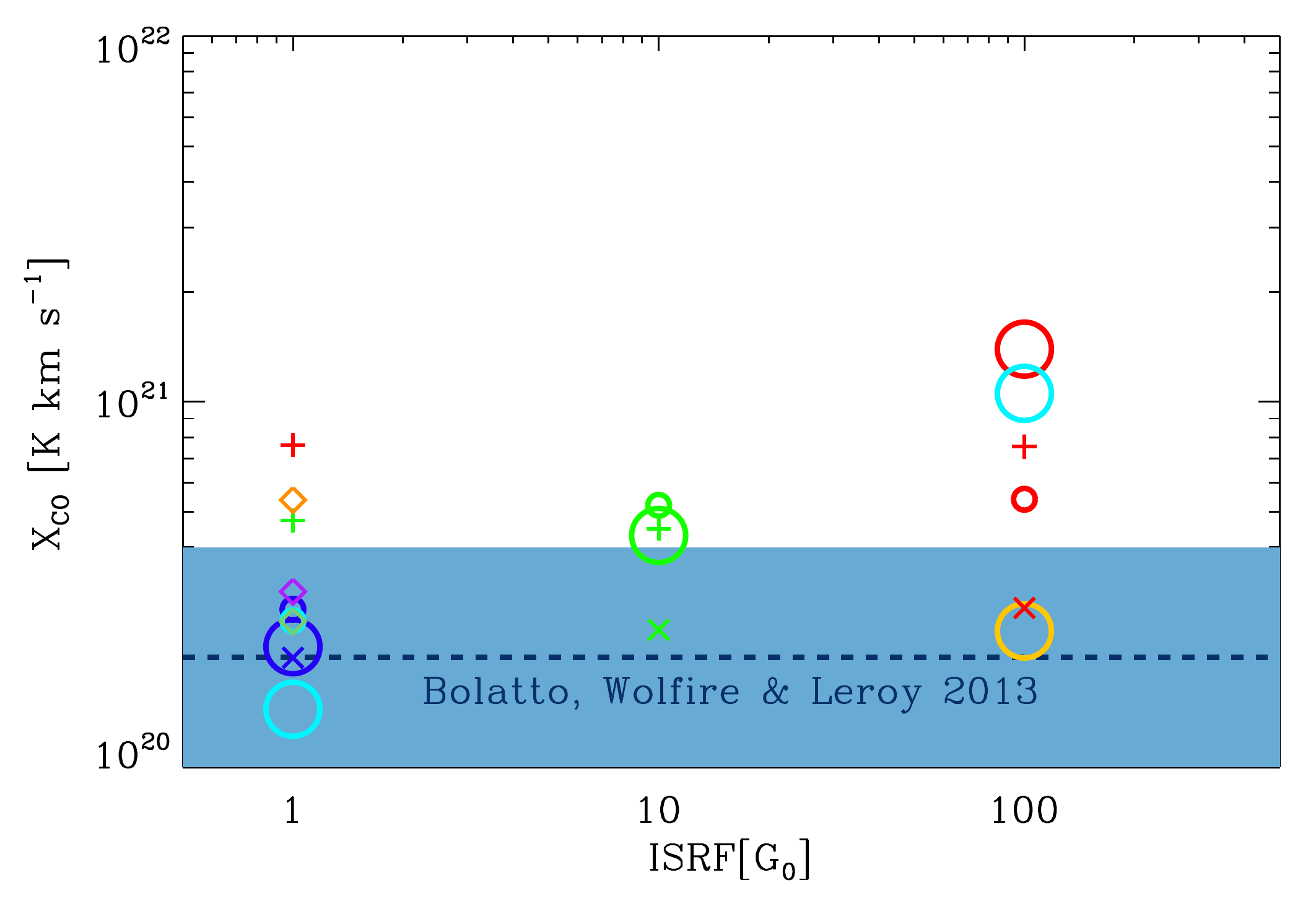}}
\caption{$X_{\rm CO}$ plotted against ISRF for all the clouds using the same labels as in Figure \ref{fig:R21vsSFR}. The line represents the typical value for $X_{\rm CO}$ and the shaded region the scatter as given by \citet{Bolatto:2013hl}. Note that clouds with metallicities of $ Z \neq Z_{\odot}$ are not included}
\label{fig:XCOvsSFR}
\end{figure}

\begin{figure}
\centering
\subfloat{\includegraphics[width=1.\linewidth]{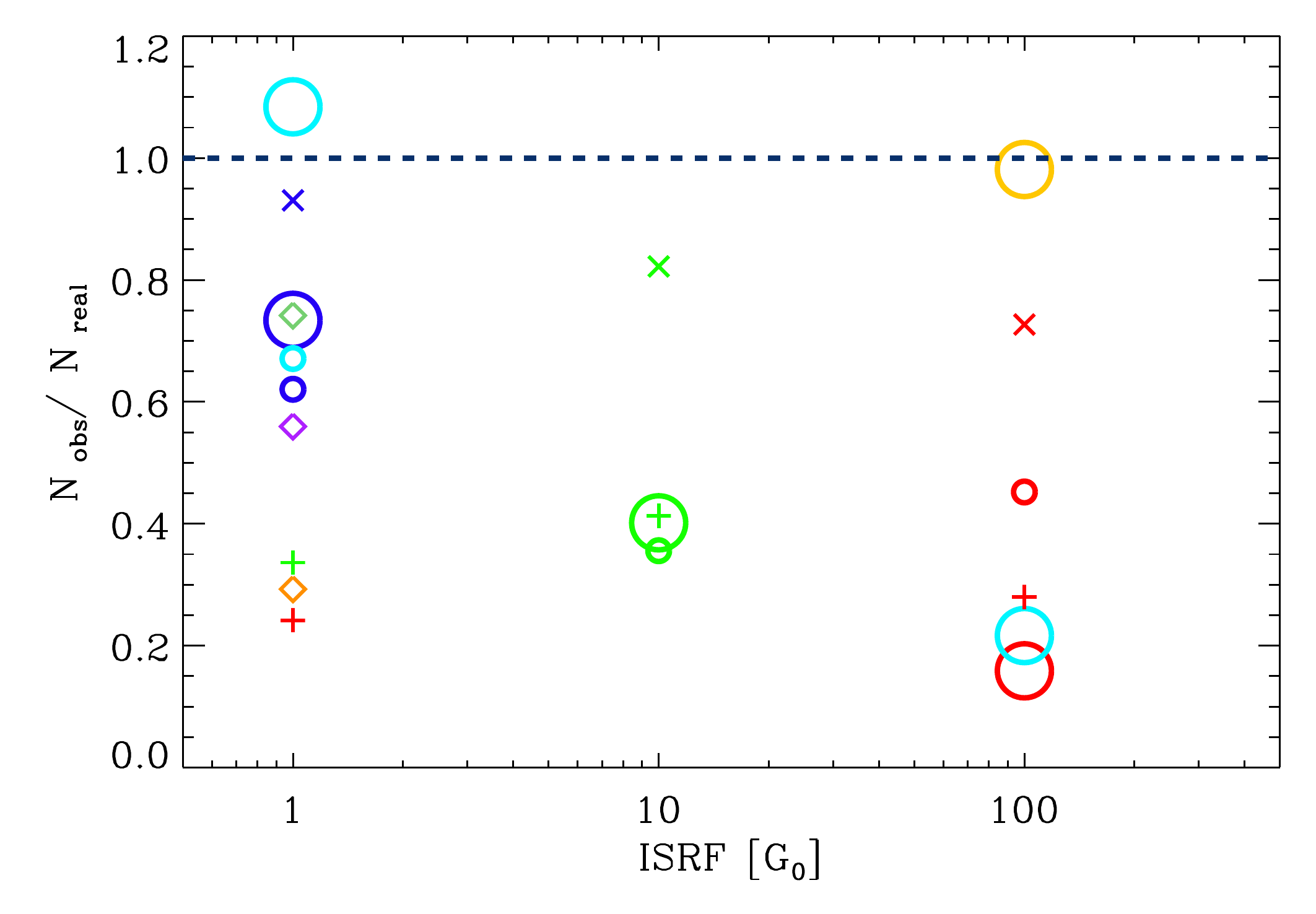}}
\caption{Shows the ratio of $N_{\rm obs}/N_{\rm real}$ against ISRF for all the clouds using the same labels as in Figure \ref{fig:R21vsSFR}, where $N_{\rm obs}$ is calculated using $R_{21}$. Note that clouds with metallicities of $ Z \neq Z_{\odot}$ are not included}
\label{fig:H2vsSFR}
\end{figure}

\begin{figure}
\centering
\subfloat{\includegraphics[width=1.\linewidth]{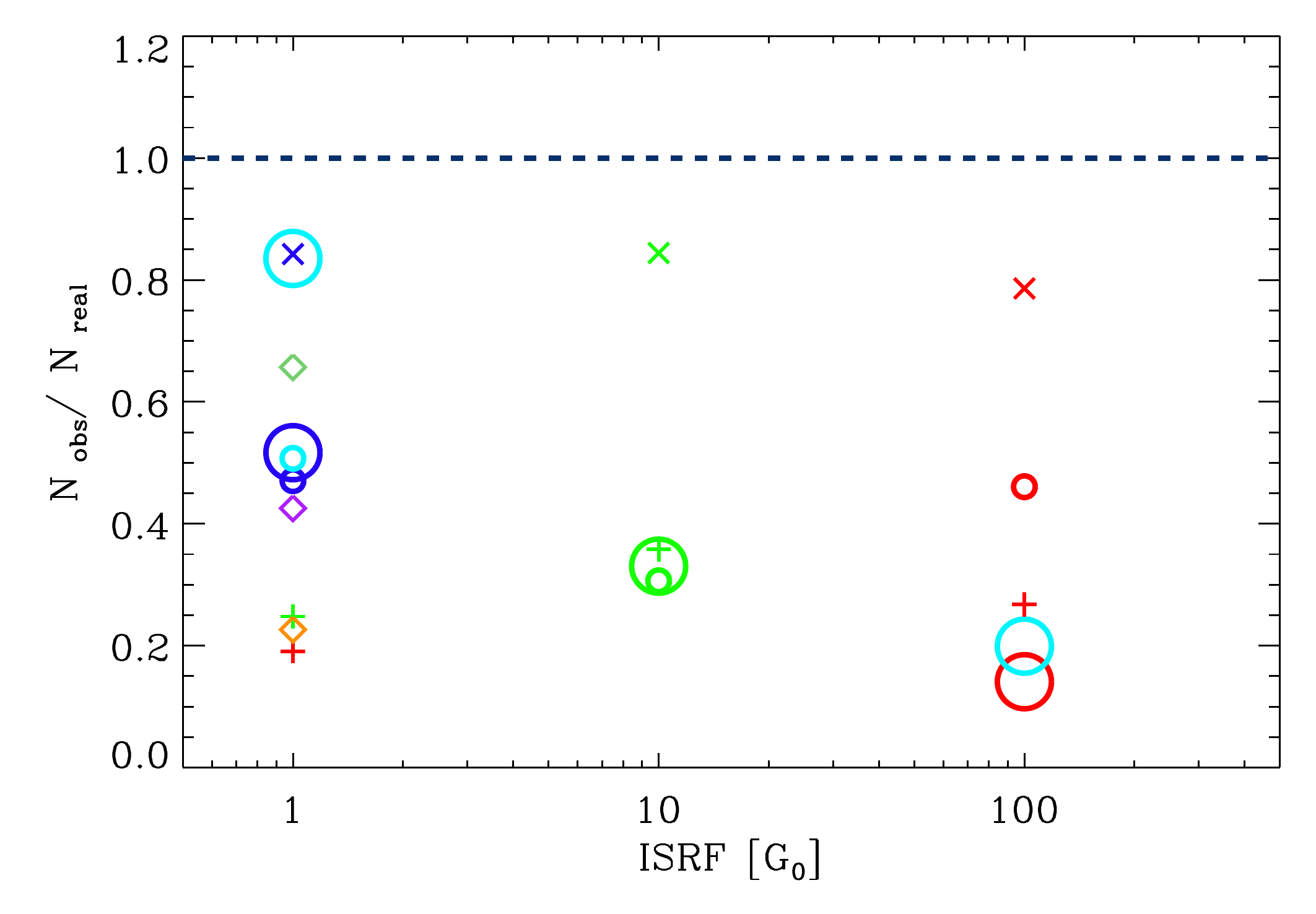}}
\caption{Similar to Figure \ref{fig:H2vsSFR} but using $R_{31}$ instead of $R_{21}$}
\label{fig:H2vsSFRR31}
\end{figure}

Before doing so we reproduce Figure 4 from \citet{2015MNRAS.452.2057C} where they plotted the value of $X_{\rm CO}$ for each cloud against the the 'star formation rate', which is a proxy for changes in the ISRF and CRIR. In this case we plot against the ISRF as well as include additional clouds that where not studied in \citet{2015MNRAS.452.2057C}(See Figure \ref{fig:XCOvsSFR}). Additionally we have only included clouds with $Z=Z_{\odot}$, since $X_{\rm CO}$ is empirically derived from observations within the Milky Way and therefore intrinsically assumes a solar-like metallicity. It is important to note that this plot may look slightly different from Figure 4 of \citet{2015MNRAS.452.2057C}. This is because our radiative transfer approach here includes the refinement routine described in \citet{2017MNRAS.465.2277P} and the Sobolev-Gnedin approximation described in Appendix \ref{appsec}. When compared, the results presented here systematically lower the value of $X_{\rm CO}$ making these results closer to the typically used value.

In Figure \ref{fig:H2vsSFR} the ratio of $N_{\rm obs}/N_{\rm real}$ against the ISRF is plotted. When calculating $N_{\rm obs}$ we have used the $X_{\rm CO} = 2 \times 10^{20}\  {\rm cm ^{-2}\ K\ km\ s}^{-1}$ as given by \citet{Bolatto:2013hl} and $R_{21} =0.7$. From this figure it becomes evident that the amount of molecular gas derived from $W_{21}$ can easily be underestimated.  This can be understood when comparing with Figure \ref{fig:XCOvsSFR} where at high ISRF the standard value of $X_{\rm CO}$ will underestimate the total column of ${\rm H_2}$. On the other hand Figure \ref{fig:R21vsSFR} shows that using an average value of $R_{21} =0.7$ will overestimate the amount of $W_{10}$ at high ISRF. Effectively, this compensates the existing biases of both conversion factors to some extent, however this is not enough to avoid underestimating the amount of ${\rm H_2}$ due to the high errors in $X_{\rm CO}$. This is also the case for the green and red plus signs at ISRF $=1$, they correspond to the high CRIR runs, where $X_{\rm CO}$ is also underestimated. 

At lower ISRF the discrepancies between $N_{\rm obs}$ and $N_{\rm real}$ arise from $R_{21}$, since at lower ISRF $X_{\rm CO}$ is well within the accepted value. From Figure \ref{fig:R21vsSFR} we can see that using $R_{21} =0.7$ effectively underestimates the amount of $W_{10}$ and therefore the total column density of ${\rm H_2}$. This effect is even stronger when using $R_{31}$ instead of $R_{21}$, which is seen in Figure \ref{fig:H2vsSFRR31} where we use $R_{31}=0.5$ to calculate $N_{\rm obs}$.  

Finally, even though our results suggest that $R_{21} =0.7$ is a good first approximation for a conversion factor between $W_{21}$ and $W_{10}$, the uncertainties and degeneracies surrounding $R_{21}$ still have a significant effect on $N_{\rm obs}$. This is also true for $R_{31}$, where the effect on $N_{\rm obs}$ is even larger given that the accepted value for $R_{31}$ is a lower limit according to our results. Alternatively, if the surrounding ISRF could be constrained -- for example, by looking at dust SEDs -- then the average values in Figure~\ref{fig:R21vsSFR} (and Figure~\ref{fig:R31vsSFR} for $R_{31}$) for each ISRF bin could be used to improve the estimate in the value of $R_{21}$ and $R_{31}$. In principle, this can lead to better true column density of molecular material. Whether this would yield a more accurate value of $N_{\rm obs}$ would depend on the uncertainties on the method used to obtain the ISRF and is beyond the scope of this paper.

\subsection{$R_{21}$ as a probe of CO abundance}

\begin{figure}
\centering
\includegraphics[width=1.\linewidth]{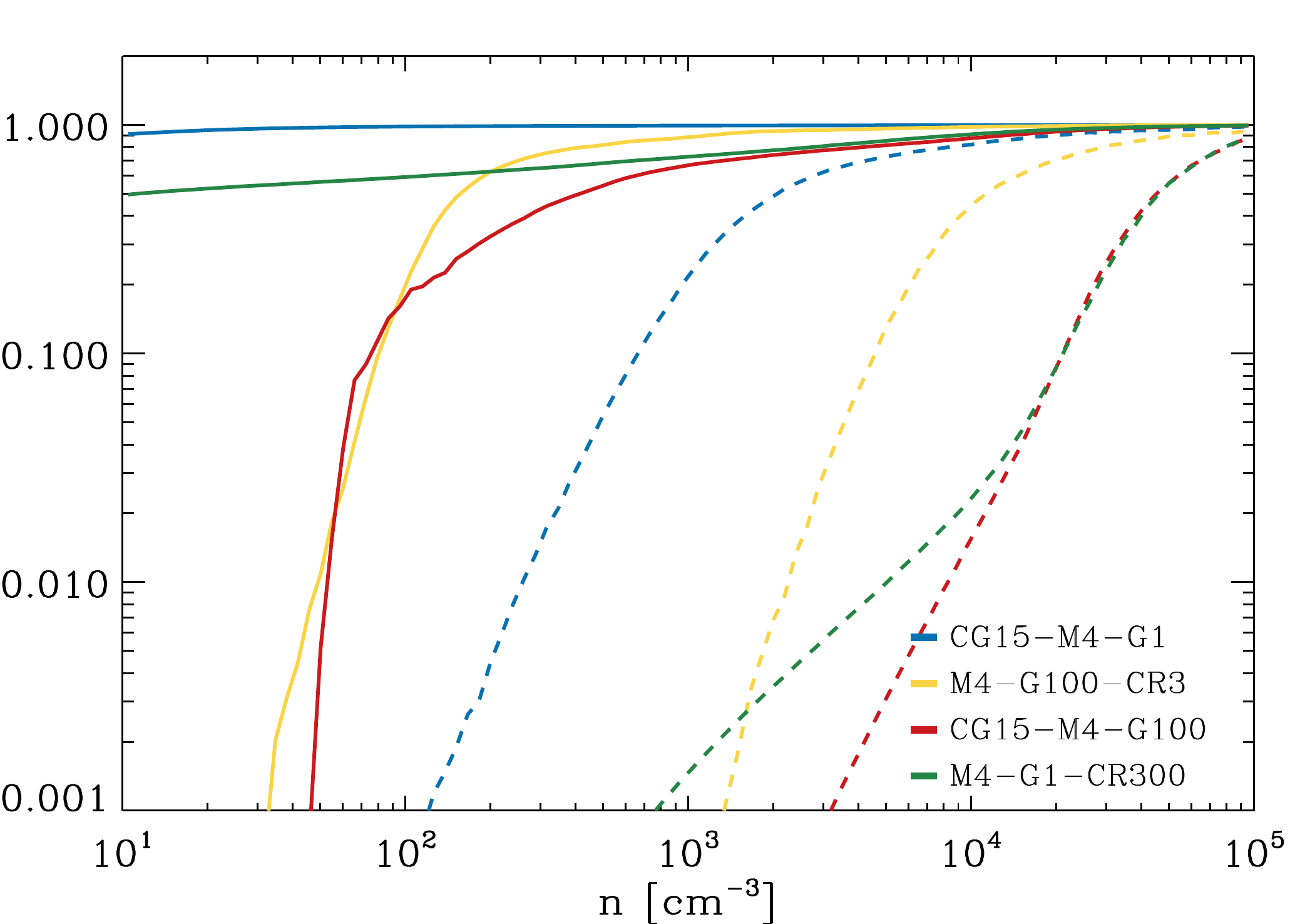}
\caption{Each coloured line represent a different simulation. Solid lines track, as a function of number density, the fraction of the mass of H in the form of H$_2$ while dashed lines track the the fraction of the mass of C in the form of CO.} 
\label{fig:XCOvsRho}
\end{figure}

As discussed in Section \ref{ressec}, variations in the strength of the ISRF and the CRIR affect $R_{21}$ in different ways. Since we are interested in quantifying the effect of the ISRF and the CRIR on the state and abundance of molecular gas we will focus on the following 4 clouds CG15-M4-G1, M4-G1-CR100, M4-G100-CR1 and CG15-M4-G100.  We then plot the fractional abundance of CO and H$_2$ as a function of the average number density of the gas for each cloud. This is shown in Figure \ref{fig:XCOvsRho}, where the solid lines represent the H$_2$ abundance fraction and the dashed line CO abundance fraction. 

First thing to note is that the H$_2$ abundance fraction only changes when the ISRF changes, while changes to the CRIR make a very small impact. Given the chemical model included in these simulations, we know that UV radiation is needed to effectively dissociate H$_2$. Even though cosmic rays can destroy some H$_2$ their overall effect is small \citep{Glover:2012et}. On the other hand cosmic rays can be very effective at reducing the total fraction of CO and can be seen when comparing the CO abundance of M4-G1-CR100 and M4-G100-CR1. The reason for this becomes evident when looking at the following chemical reactions 

\begin{align*}
{\rm He} + {\rm c.r} \rightarrow {\rm He^+} + {e^-},\\
{\rm CO} + {\rm He^+} \rightarrow {\rm C^+} + {\rm O} + { \rm He}.
\end{align*}

When looking at the cloud with high ISRF the fraction of CO abundance (dashed yellow line) increases quickly, this is because once the CO is well shielded the production of CO is efficient. As a result the emission coming from these high-density regions will be bright and well correlated with high values of $R_{21}$. On the other hand when looking at the high CRIR cloud, the fraction of CO abundance (dashed green line) starts increasing at similar number densities ($\sim 10^3 \ {\rm cm^{-3}}$), however, at a much slower rate. This is because the CO production is being constantly hampered by the cosmic rays which are not attenuated. As such the emission from these regions will be faint due to the low abundance of CO, more importantly $R_{21}$ will have values around $\sim 0.3$. In addition, the increase in the CRIR will heat up the gas and therefore slightly increase the average value of $R_{21}$ (See Figure~\ref{fig:R21vsSFR}).

A recent paper by \citet{2017ApJ...839...90B} studied how increasing the CRIR can be important in destroying CO. In Figure 11 of this paper they compare the CO/H$_2$ fraction as a function of number density, for varying CRIR. We reproduce this figure with our own set of simulations for which the CRIR is increased in the same way (Note to better compare to their results we ran an additional simulation with $\zeta_{\rm H} = 3 \times 10^{-14}\ {\rm s^{-1}}$ that was not included in our initial setup). Our results show a similar trend where the CO/H$_2$ abundance ratio decreases with increasing CRIR. 

\begin{figure}
\centering
\includegraphics[width=1.\linewidth]{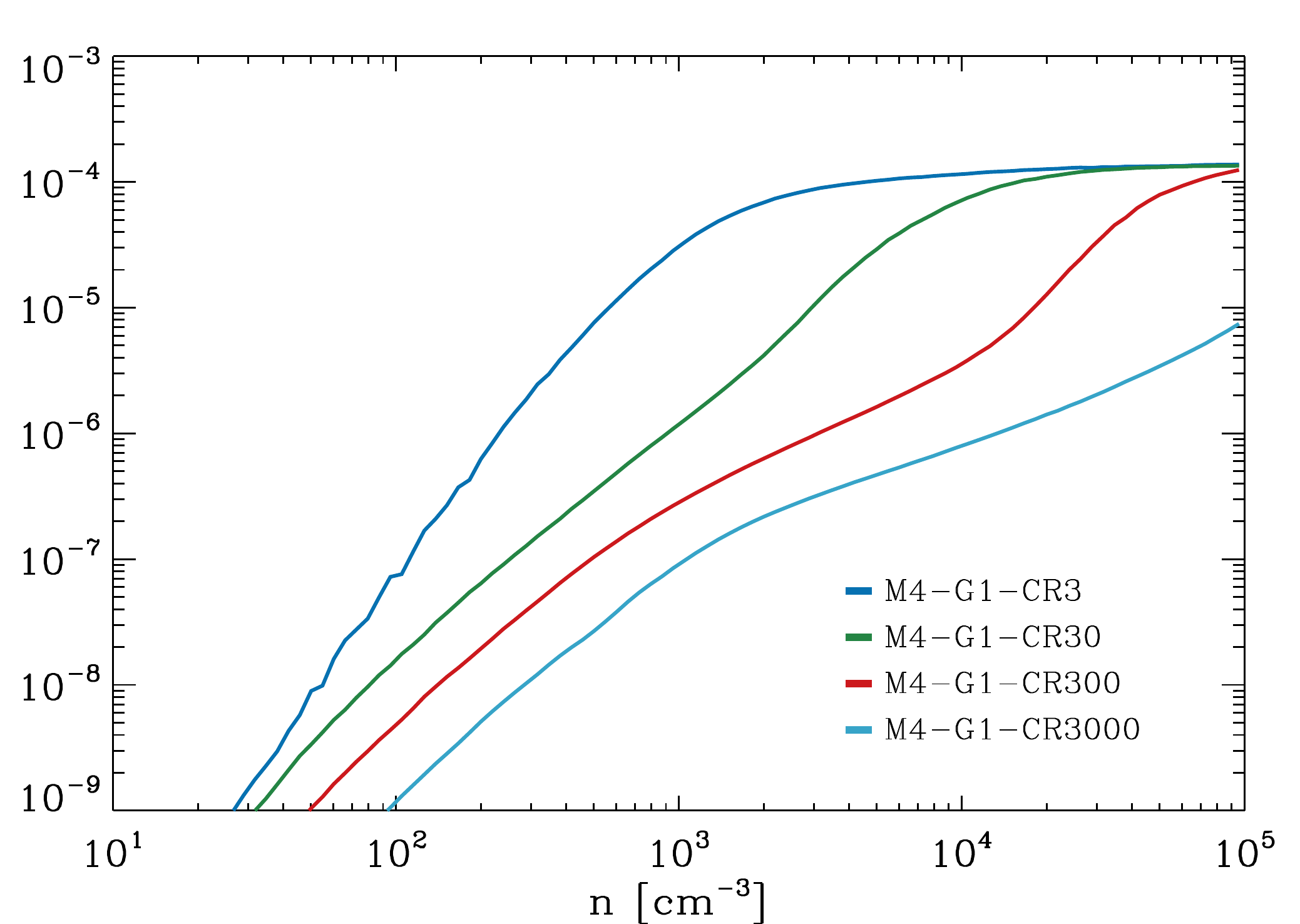}
\caption{Illustrates how CO/H$_2$ abundance ratio changes with average number density.} 
\label{fig:CO_H2vsRho}
\end{figure}

The effect of CRIR on the abundance of CO will have a direct impact on the CO emission and therefore how much molecular gas can be traced within GMCs. Even though the total CO emission is reduced, our synthetic observations show that the changes in abundances seen above can be traced to some extent when looking at the resolved integrated intensities and $R_{21}$ of these clouds (see Figure \ref{fig:WPDFvsR21}). One caveat to keep in mind when considering these results, is that our models have a constant CRIR throughout the whole cloud -- cosmic rays are not attenuated and therefore able to reach the densest regions of the cloud. Whether this is an accurate approximation is beyond the scope of this study and therefore the reader should keep this in mind when interpreting these results. 

\section{Conclusions}\label{consec}

We have studied a range of numerical models of molecular clouds, in which the initial cloud parameters, and properties of the ISM, were systematically varied. We performed radiative transfer calculations on the simulations just before the onset of star formation in each case, to create synthetic CO line emission cubes for all clouds. We used these synthetic observations to study the impact of environment on CO line emission and CO line ratios. Our main findings can be summarised as follows:

\begin{enumerate}
\item The value of $R_{21}$ and its correlation with dense/cold and warm/diffuse, allows it to act as a probe of the conditions of the gas within GMCs.  From all the environmental changes studied, variations in the ISRF and the CRIR have the largest impact on the average value and the cumulative PDF of $R_{21}$. 
\item The dependence of different line-ratios ($R_{21}$, $R_{32}$ and $R_{31}$) on environment can also be observed when looking at unresolved clouds where the total emission is averaged. Our results suggest that the accepted values for $R_{21}$ and $R_{31}$ are a good first approximation. At the same time the scatter around the accepted value ($ \sim \pm 0.2 $) suggests that careful consideration should be had when using them as conversion factors, specially given the high dependence on the ISRF. 
\item When calculating the column density of H$_2$ molecular gas of GMCs it is important to consider the biases of $X_{\rm CO}$ and line ratios $R_{21}$ and $R_{31}$. At a high ISRF ($G_0 = 100$) $X_{\rm CO}$ will underestimate $N_{\rm H_2}$. This is only slightly compensated by the bias line ratios have at high ISRF. On the other hand since at low ISRF ($G_0 = 1$) $X_{\rm CO}$ is well constrained, the errors in $N_{\rm H_2}$ come from line ratios underestimating the total amount of emission from the $J=1-0$ transition line. 
\item Cosmic rays can help regulate the total CO abundance within GMCs. When $\zeta_{\rm H} = 3 \times 10^{-17}\ {\rm s^{-1}}$ the CO to H$_2$ abundance ratio is $\sim 10^{-4}$ at densities of $\sim 10^3 {\rm cm^{-3}}$. As $\zeta_{\rm H}$ is increased the CO to H$_2$ fraction is considerably reduced reaching values of only $\sim 10^{-5}$ at densities of $\sim 10^5 {\rm cm^{-3}}$ for rates of $\zeta_{\rm H} = 3 \times 10^{-14}\ {\rm s^{-1}}$. This has a direct impact on the CO emission and on the average value and distribution of $R_{21}$ 
\end{enumerate}

\section*{Acknowledgements}
PCC acknowledges support from the Science and Technology Facilities Council (under grant ST/N00706/1) and the European Community's Horizon 2020 Programme H2020-COMPET-2015, through the StarFormMapper project (number 687528).
SCOG and RSK acknowledge financial support from the Deutsche Forschungsgemeinschaft  via SFB 881, ``The Milky Way System'' (sub-projects B1, B2 and B8) and SPP 1573, ``Physics of the Interstellar Medium''. They also acknowledge support from the European Research Council under the European Community's Seventh Framework Programme (FP7/2007-2013) via the ERC Advanced Grant STARLIGHT (project number 339177).

\bibliography{Bibliography}
\bibliographystyle{mn2e}

\appendix
\section{Sobolev-Gnedin approximation}\label{appsec}

The Sobolev approximation, most commonly known as the Large Velocity Gradient (LVG) approximation, is widely used when calculating the level populations of a gas within GMCs. In the original paper, \citet{1957SvA.....1..678S} studied the idea that emission coming from gas within a moving medium will be doppler-shifted before it can be reabsorbed and therefore escapes. Whether a photon escapes or not is determined by the velocity gradient ($|\nabla \vec{v}|$), i.e. the larger $|\nabla \vec{v}|$ is the smaller is the volume where the photon can be reabsorbed. RADMC-3D already makes use of the Sobolev approximation when calculating level populations, the detailed implementation can be found in \citet{Shetty:2011eh}. 

Even though \citet{1997NewA....2..365O} showed that the Sobolev approximation is a robust approximation for GMCs, the method is still limited by the fact that it only considers changes in velocity. When using the Sobolev approximation the aim is to calculate the probability that a photon will escape a given region. The escape probability is given by: 

\begin{equation}
\beta=\frac{1-e^{-\tau_{v}}}{\tau_{v}},
\label{eqn:EscProb}
\end{equation}

\noindent where $\tau_v$ is the optical depth given a velocity gradient. $\tau_v$ can be calculated by

\begin{equation}
\tau_v = \frac{c^3}{8\pi \nu^3_{ij}} \frac{A_{ij} n}{1.064|\nabla \vec{v}|} \left( f_j \frac{g_i}{g_j} - f_i \right),
\label{eq:tau_V}
\end{equation}

\noindent where $\nu_{ij}$ is the emission frequency of the transition, $g_i$ and $g_j$ are the statistical weights for each level, $n$ the total number density and, $f_i$ and $f_j$ is the fractional population level \citep{2007A&A...468..627V}. 

Now consider a scenario in which $|\nabla \vec{v}|$ is small but the change in density i.e. the density gradient ($|\nabla \rho|$) is high. According to Equation \ref{eq:tau_V} $\tau$ would be high and therefore the probability of a photon escaping would be low. However given that $|\nabla \rho|$ is higher the escape probability of the photon should be higher, yet Equation \ref{eq:tau_V} does not consider $|\nabla \rho|$ and therefore the escape probability is unchanged. For the GMCs considered in this study we can foresee different cloud regions where this scenario is likely: dense cores can have low velocity dispersions, yet can have very steep density gradients. In effect, the LVG is a local approximation, in that it constructs a length-scale based on local properties. These local properties, such as density, temperature, etc, are then assumed to be held constant over this length-scale, allowing us to derive an optical depth.  If, however, the density varies rapidly, then the LVG approximation makes an error.  Thankfully, it is relatively simple to improve this approximation. 

\subsection*{Gnedin Length Approximation}

The Sobolev approximation is a local approximation given a certain length scale, where the length scale is calculated from the velocity gradient. Commonly this is called the Sobolev length and is defined by:

\begin{equation}
L_{Sob} = \frac{v_{th}}{|\nabla \vec{v}|},
\end{equation}  

\noindent where $v_{th}$ is the thermal velocity. From this and equation \ref{eq:tau_V} one can reconstruct a more general solution for the opacity where $L$ need not be $L_{sob}$ but any other length scale relevant to the problem at hand. One such length scale is the one presented by \citet{2009ApJ...697...55G} where they define a length scale base on density gradients in order to determine column densities of H$_2$. They define this length scale as

\begin{equation}
L_{Gn} = \frac{n_{\rm H_2}}{|\nabla n_{\rm H_2}|},
\end{equation}  

\noindent where $n_{\rm H_2}$ is the number density of H$_2$, however for our purposes we may define this in terms of the number density of the molecule that is being modelled. With this in mind we can rewrite equation \ref{eq:tau_V}, to calculate an optical depth ($\tau_{\rho}$) given a density gradient by

\begin{equation}
\tau_{\rho} = \frac{h c}{4\pi \sqrt{\pi} v_{th}} \frac{n^2}{|\nabla n |} \left( f_j \frac{g_i}{g_j} - f_i \right),
\label{eq:tau_rho}
\end{equation}

\noindent that has the same form as Equation \ref{eq:tau_V}. At the same time Equation \ref{eq:tau_rho} would fail to accurately calculate the escape probability given that $\tau_{\rho}$ does not consider the gradient in velocity, especially in highly turbulent systems such as GMCs.  

At this point we have two adequate yet limited ways of calculating an optical depth. To obtain the best from each we follow the approach taken by \citet{2015ApJ...799..114H}, where they take the harmonic mean between $\tau_{v}$ and $\tau_{\rho}$, as follows

\begin{equation}
\tau = \frac{\tau_v \tau_{\rho}}{\tau_v  +  \tau_{\rho} }.
\end{equation}

The resulting optical depth will then be used in equation \ref{eqn:EscProb} to calculate the escape probability. 

\subsection*{Implementation and Testing}

\begin{figure}
\centering
\includegraphics[width=1.\linewidth]{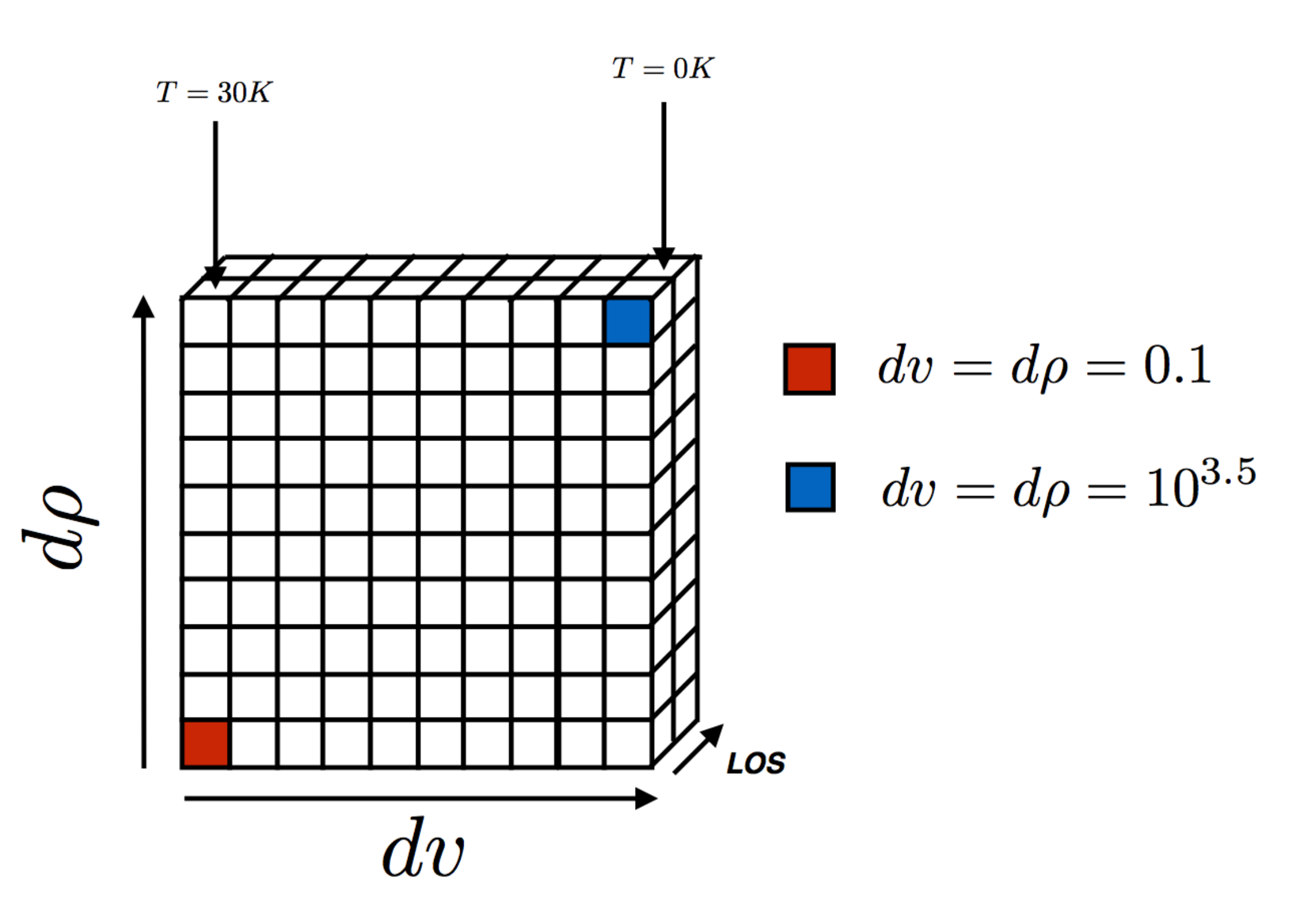}
\caption{The setup of the grid used as an input for the radiative transfer. All the cells in the the front slice have the same density, temperature and velocity. The back slice has a temperature of $0$ K. Densities and velocities take a value according to $dv$ or $d\rho$, which are increased such that $dv=d\rho$ over the diagonal. $dv$ is increased accordingly over the $x$-axis and $d\rho$ over the $y$-axis. } 
\label{fig:LVGsetup}
\end{figure}

We have used the underlying framework of RADMC-3D and implemented the Gnedin approximation, as described above, as an improvement to the already present Sobolev approximation. In order to test this method we set up a test scenario that highlights the difference between LTE, LVG and LVG+ (Sobolev-Gnedin approximation). 

We set up an input grid of 10 by 10 by 2, Figure \ref{fig:LVGsetup} shows the position-position-position (PPP) cube that was used as an input into RADMC-3D. All the cells in the front slice have the same physical properties and will therefore have the same intrinsic emission in a scenario where $|\nabla \vec{v}|$ or $|\nabla \vec{\rho}|$ are not considered. The back slice serves to calculate the velocity and density gradients used to obtain the optical depth of each cell in the front slice. To keep the setup simple, we increase $|\nabla \vec{v}|$ only in the $x$-axis while keeping $|\nabla \vec{\rho}|$ constant. Conversely we increase $|\nabla \vec{\rho}|$ while keeping $|\nabla \vec{v}|$ constant on the $y$-axis. The increase in the gradient is taken with respect to the values of density and velocity in the front slice. These are such that $dv=d\rho$ will be the same whenever $x=y$. Additionally to avoid confusion we set the temperature of each cell in the back slice to be $T = 0$ K so that these cells have no emission.  

\begin{figure}
\centering
\includegraphics[width=1.\linewidth]{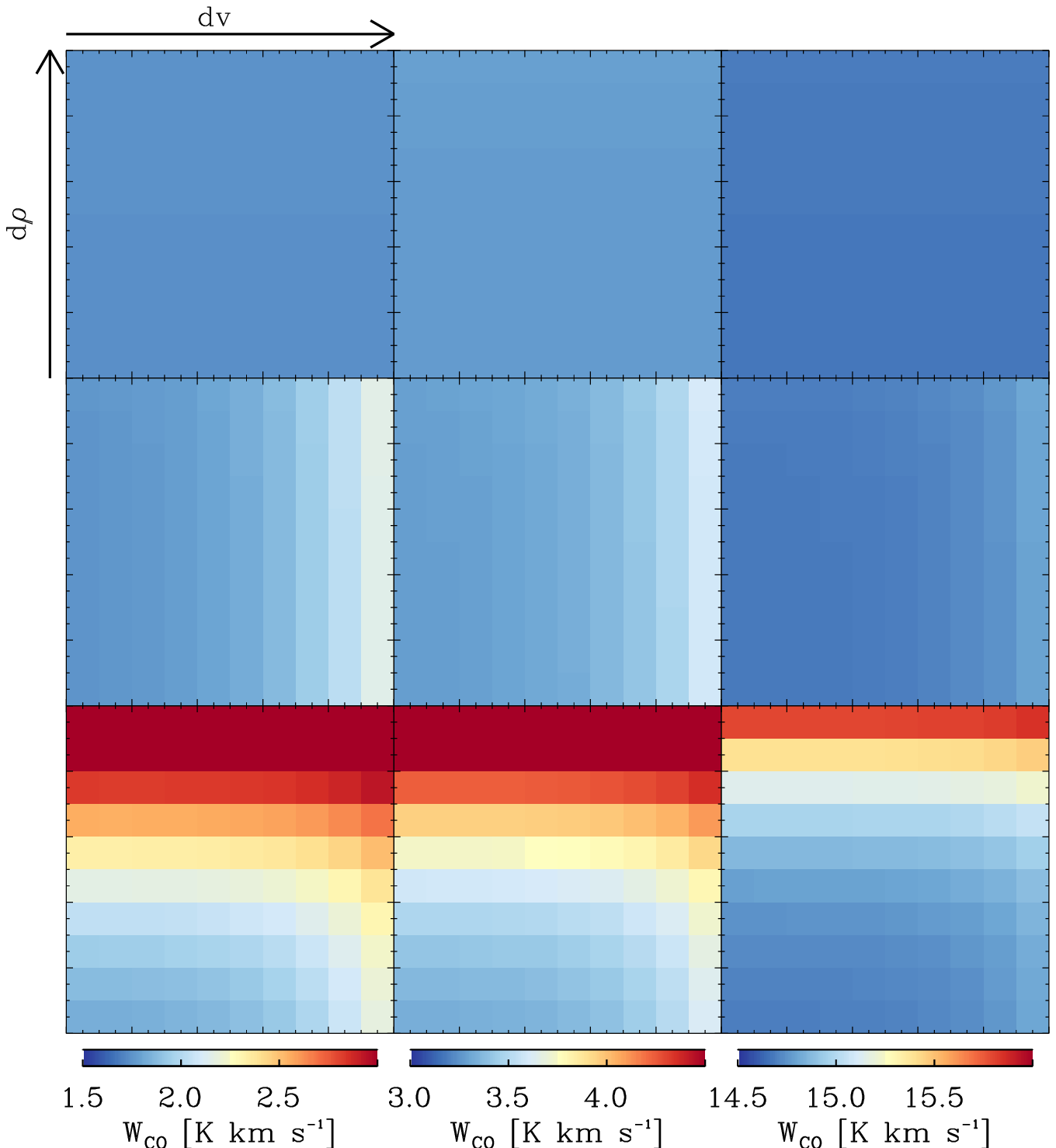}
\caption{The integrated intensities of CO $J=1-0$ for each run on each cube. Each row has a different method for calculating level populations: top  = LTE, middle = LVG, bottom  =LVG+. Similarly each column has a different number density for the front slice, left: $n=50\ {\rm cm}^{-3}$, middle: $n=100\ {\rm cm}^{-3}$, right: $n=500\ {\rm cm}^{-3}$. Note that even though the colours are similar the colourbar for each column is different} 
\label{fig:LVGplustest}
\end{figure}

Given the purpose of this paper, we perform the radiative transfer for CO $J=1-0$ but in principle the method works for any other molecule or line. Additionally we create 3 different cubes for 3 different densities of the front slice ($n=50\ {\rm cm}^{-3}$, $n=100\ {\rm cm}^{-3}$ and $n=500\ {\rm cm}^{-3}$). We run 3 radiative transfer simulations for each cube, each of these uses either LTE, LVG or LVG+. The resulting integrated intensities for each run for each cube are shown in Figure \ref{fig:LVGplustest}.

As expected changing the method for calculating the population levels makes a big difference in the final image. Since the density and temperature of each cell are exactly the same assuming LTE results in the integrated intensity for each cell also being the same. This is because LTE uses no information of the gradient surrounding the cell to calculate the level populations. For the LVG scenario we can definitely see a change in intensity over the $x$-axis, which corresponds to an increase in $|\nabla \vec{v}|$. However not taking $|\nabla \rho|$ into account results in the intensity being exactly the same on the $y$-axis. Finally when using the LVG+ method both changes on $|\nabla \vec{v}|$ and $|\nabla \rho|$ are reflected on the final intensity of each cell. 

One important note to keep in mind when looking at these results is why increasing the gradient results in an increase in intensity rather than a decrease. The reason is that the background radiation and the high temperature of the cell causes higher levels to be radiatively pumped. These will then quickly cascade down to the the ground state i.e $J=1-0$ causing it to be much brighter. The reason behind this is that for lower temperatures the difference between cells is very small and hard to see. This should not be a problem since our interest is to test that the method works. In the next Section we will show how using LVG+ changes the integrated intensities of one of our clouds.

\subsection*{Result Comparison}

\begin{figure*}
\centering
\includegraphics[width=1.\linewidth]{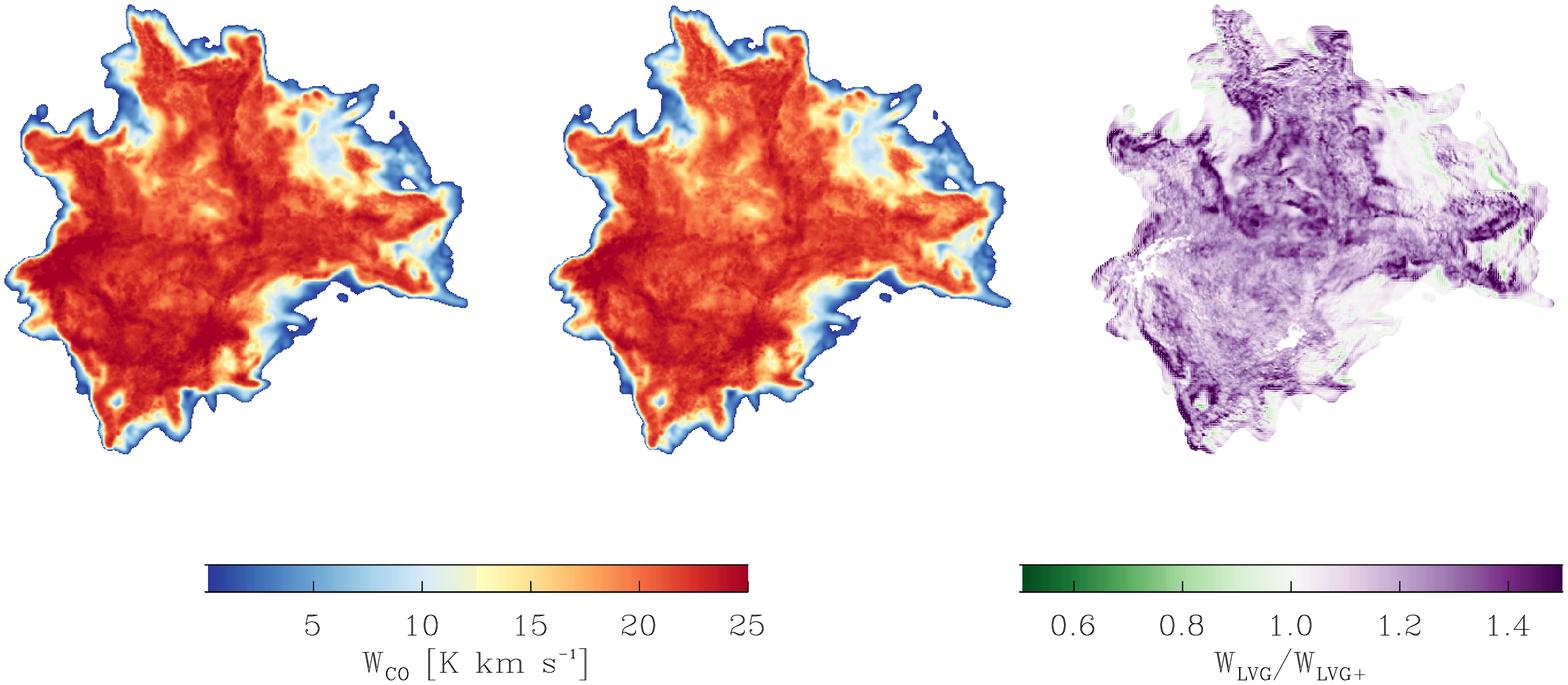}
\caption{The image on the left shows the integrated intensity for CG15-M4-G1 with an LVG treatment while the middle image does so for LVG+. The image on the right is the ratio of LVG/LVG+ integrated intensities } 
\label{fig:diffLVG}
\end{figure*}

Even though we tested that LVG+ works, it is important to check whether this new method has any significant effect on a more realistic system such as our numerically simulated clouds. To check this we performed two radiative transfer simulations on CG15-M4-G1, one using LVG and another using LVG+. The integrated intensities of these two runs are shown in Figure \ref{fig:diffLVG}.

At first glance it seems that the difference between LVG and LVG+ is negligible. However when taking the ratio between the both we can see that the brightness of the cloud definitely changes when using LVG+. For most regions of the cloud the brightness is reduced, nonetheless there are still regions that can either become brighter or remain the same. It then follows that considering changes in the local density does play an important role when calculating the level populations and total emission of a system. This illustrates how small changes in the physical conditions of the gas and the interplay with the surrounding environment can lead to changes in the total emission.

\end{document}